\documentclass[epj,final]{svjour}
\usepackage{latexsym}
\usepackage{graphicx}
\usepackage{latexsym,setspace}
\usepackage{amsmath,amssymb,graphicx}
\usepackage{color,float,placeins}
\usepackage{colortbl,array,multirow}
\usepackage{ulem}

\begin{document}

\title{Dynamo regimes and transitions in the VKS experiment}

\author{M. Berhanu\inst{1} \and G. Verhille\inst{2} \and J. Boisson\inst{4} \and B. Gallet\inst{1} \and C. Gissinger\inst{1} \and S. Fauve\inst{1} \and  N. Mordant\inst{1} \and F. P\'etr\'elis\inst{1} \and M. Bourgoin\inst{3} \and Ph. Odier\inst{2} \and J.-F. Pinton\inst{2} \and N. Plihon\inst{2} \and S. Auma\^\i tre\inst{4} \and A. Chiffaudel\inst{4} \and F. Daviaud\inst{4} \and B. Dubrulle\inst{4} \and C. Pirat\inst{4}}
\institute{Laboratoire de Physique Statistique, CNRS \& \'Ecole Normale Sup\'erieure,\\ 24 rue Lhomond, F-75005 Paris, France \\
\and Laboratoire de Physique de l'\'Ecole Normale Sup\'erieure de Lyon,  CNRS \& Universit\'e de Lyon,\\ 46 all\'ee d'Italie, F-69364 Lyon cedex 07, France\\
\and Laboratoire des \'Ecoulements G\'eophysiques et Industriels, CNRS \& Universit\'e Joseph Fourier,\\ BP53, F-38041, Grenoble cedex 9, France\\
\and Service de Physique de l'\'Etat Condens\'e, CNRS \& CEA Saclay, F-91191 Gif-sur-Yvette, France}

\authorrunning{M. Berhanu et al}

%\titlerunning{M. Berhanu et al.}

\date{\today}

\abstract{ 
The Von K\'arm\'an Sodium experiment yields a variety of dynamo regimes, when asymmetry is imparted to the flow by rotating impellers at different speed $F_1$ and $F_2$. We show that as the intensity of forcing, measured as $F_1+F_2$, is increased, the transition to a self-sustained magnetic field is always observed via a supercritical bifurcation to a stationary state. For some values of the asymmetry parameter $\theta=(F_1-F_2)/(F_1+F_2)$, time dependent dynamo regimes develop. They are observed either when the forcing is increased for a given value of asymmetry, or when the amount of asymmetry is varied at sufficiently high forcing. Two qualitatively different transitions between oscillatory and stationary regimes are reported, involving or not a strong divergence of the period of oscillations. These transitions can be {interpreted} using a low dimensional model based on the interactions of two dynamo modes.
}

\maketitle

\section{Introduction}
%%%%%%%%%%%
While still being quite far from the parameter regime that characterizes natural objects, dynamo experiments using liquid metals have the advantage of having adjustable control parameters. They also display dynamical evolutions that can be recorded over long durations compared to the Joule characteristic time. The Riga~\cite{Gailitis:PRL:01} and Karlsruhe~\cite{Stieglitz:PoF:01} experiments have established the central role of helicity and shear in the dynamo process, with dynamo characteristics well predicted by laminar models -- although the underlying flows are turbulent, with a moderate turbulence rate. In the recent VKS experiment~(see \cite{Monchaux:PoF:09} and references therein), the situation is different since non axisymmetric velocity components (in the bulk and near the impellers) play a leading role in the magnetic field generation: the observed axisymmetric dynamo mean field cannot be generated by the axisymmetric mean flow alone.  Another central observation in the VKS experiment lies in the variety of dynamo regimes observed when the flow driving parameters and magnetic Reynolds number are varied~\cite{Monchaux:PoF:09}.  We report here on the bifurcations observed in the VKS dynamo based on a thorough study of the parameter space. The von K\'arm\'an flow of sodium is generated inside a cylinder by the counter rotation of coaxial impellers at frequencies $(F_1, F_2)$ (see Fig.~\ref{fig1}a). When both impellers rotate at the same frequency $F_1 =F_2$, the driving and hence the mean flow structure are symmetric with respect to any rotation ${\cal R}_\pi$ of $\pi$ around any radial axis in its equatorial plane. When the frequencies $F_1$ and $F_2$ are different, this symmetry is broken. One possible variable to quantify this asymmetry is the parameter $\theta = (F_1- F_2)/(F_1+F_2)$. In addition, the choice of frequencies $(F_1,F_2)$ imposes a mean shear $F = (F_1+F_2)/2$. When the parameters $(F, \theta)$ are varied, various types of dynamos are observed. We concentrate on the following issues: \\
(i) characteristics of the bifurcation to a dynamo regime when $F$ increases, for a fixed value of the asymmetry parameter $\theta$. Our observation is that a (statistically) steady dynamo is always generated first, via a supercritical bifurcation. Time-dependent regimes can develop as a secondary bifurcation when $F$ is increased further for particular intervals of $\theta$.\\
(ii) Transition between dynamo regimes (above critical), in particular changes from stationary to oscillatory dynamics when $\theta$ is varied at constant $F$. 

In some cases, we observe a divergence of the period of oscillation during transition or bifurcations. In other cases the period remains finite.\\

The next section describes the experimental set-up and gives a summary of the dynamo capacity of the various flows
configurations studied so far in the VKS experiment. The parameter space and the bifurcations observed when increasing the forcing at a given value of asymmetry are presented in section 3. Section 4 describes how the dynamo undergoes transitions between various regimes as the asymmetry is varied at a given forcing. In section 5, we show that these observations can be understood using the predictions of a low dimensional model, involving the interactions of two dynamo modes. A final discussion is given in section 6.

\section{Experimental set-up and configurations}
%%%%%%%%%%%

\subsection{Set-up}\label{setup}

\begin{figure}
\centerline{\includegraphics[width=8.5cm]{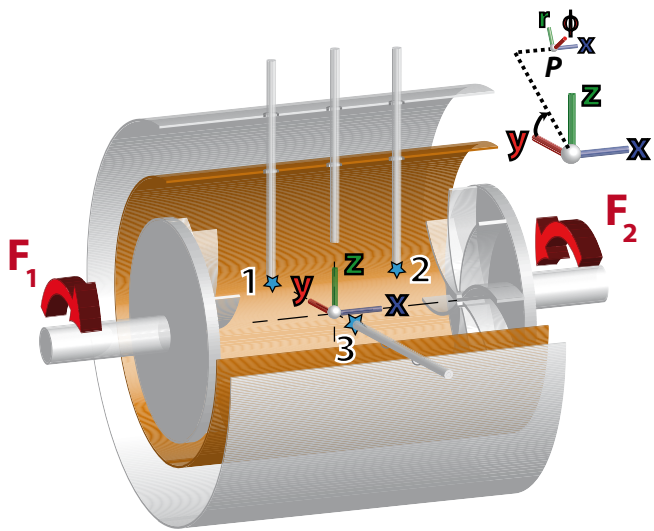}}
\caption{Experimental setup, showing the location of the Hall probes. $x$ is the axial coordinate directed from impeller 1 to impeller 2, $r$ and $\phi$ the radial and azimuthal coordinates.}
\label{fig1}
\end{figure}

The present set-up is displayed in Fig.\ref{fig1}. A von K\'arm\'an swirling flow is generated in a cylindrical vessel (radius $R_{\rm vessel}$=289~mm, length $L$=604~mm) by two counter-rota\-ting impellers 371~mm apart. The flow is surrounded by sodium at rest enclosed between the copper outer vessel and an inner copper cylinder (radius $R_{\rm cyl}=206$~mm, length $H=524$~mm). The impellers are made of soft iron disks (radius $R_{\rm imp}=154.5$~mm) fitted with 8 curved blades with height $h = 41.2$~mm. Their rotation rate can be adjusted independently to $(F_1, F_2)$. The arrows in Fig.~\ref{fig1} define the positive rotation rate $F_1,F_2 > 0$. It corresponds to the case where the curved blades are counter-rotating and ``unscooping" the fluid (pushing the fluid with the convex side of the blades). The fluid is liquid sodium (density $\rho=930 \; {\rm kg}\cdot{\rm m}^{-3}$, electrical conductivity $\sigma = 9.6\cdot10^{6} \ {\Omega}\cdot{\rm m}^{-1}$, kinematic viscosity $\nu = 6.7\cdot10^{-7} \; {\rm m}^{2}\cdot{\rm s}^{-1}$, at 120$^{\rm o}{\rm C}$). The driving motor power is 300 kW and cooling by an oil circulation inside the wall of the outer copper vessel allows experimental operation at constant temperature in the range $110 - 140^{\circ}$C. This set-up is a slightly modified version of the one previously described in~\cite{Monchaux:PoF:09}: the inner copper ring {that was located in the mid-plane} has been removed. A hydrodynamic study of this configuration has been done in \cite{Ravelet:PoF:2005} and the influence of the inner ring on the flow has been studied in details in~\cite{Cortet:PoF:09}. 

The magnetic field is measured with Hall probes inserted inside the fluid, as shown in Fig.\ref{fig1}. Unless otherwise stated, the measurements presented in this paper were made using the probe at location 1. Because of the small number of probes that were available for the experimental data studied here, no statement will be made on the spatial distribution of the observed dynamo modes. They will only be distinguished by their amplitude and time dynamics at the point of measurement. 

In the following subsection, we give a brief summary of the main results obtained with the different configurations studied so far in VKS: the inner copper wall and ring can be inserted or removed and the impellers material can be varied independently. We define the kinetic Reynolds number of the flow: {$Re =  2\pi R_{\rm imp}FR/\nu$ with $R=R_{\rm cyl}$ if the inner cylinder is present and $R=R_{\rm vessel}$ otherwise.} The corresponding magnetic Reynolds number is defined in the same way: ${R_m} = 2\pi R_{\rm imp}FR\mu_0\sigma$, where $\mu_0$ is the magnetic permeability of vacuum. {This definition is different from the one used in previous publications (it leads to $25\%$ higher values for $R_m$) but it was chosen here because it contains explicitly the flow scale $R$ and thus allows quantitative comparisons between cases with and without the inner cylinder. Since the asymmetry between the rotation rates of the impellers is a key ingredient in the results presented here, we also define individual magnetic Reynolds numbers based on the velocity of each impeller: ${R_m}_{1,2} = 2\pi R_{\rm imp}F_{1,2}R\mu_0\sigma$.} Finally, note that the conductivity of sodium is quite sensitive to temperature variations in the vicinity of its melting point ($\pm 2\%$ on $R_m$ for a temperature variation of $\pm 6^{\rm o}$C around 125$^{\rm o}$C). These variations are taken into account in the computation of ${R_m}$.

\begin{table*}
\begin{center}
\begin{spacing}{0.9}
\begin{tabular}{| c || l | l | l |} 
\hline
 &  ~inner cylinder+ring  &  ~inner cylinder alone  & ~no inner cylinder, no ring \\ 
impeller 1/impeller 2 &  (${R_m}=2.42F$ @ 120${\rm ^{\circ}C}$) &  (${R_m}=2.42F$) & (${R_m}=3.40F$)\\ 
\hline
\hline
SS/SS & no dyn. @ $F < 29$ Hz & no dyn. @ $F< 28$ Hz & no dyn. @ $F< 24$ Hz  \\
\hline
%\xout{\;\;\;\;\;\;\;\;\;\;\;\;\;\;\;\;\;\;\;\;\;\;\;\;\;\;\;\;\;\;\;\;\;\;\;}
\multirow{4}{*}{Iron/SS }& \cellcolor[gray]{0.8} &  \cellcolor[gray]{0.8} & A: $F^c=15$ Hz\\
 & \cellcolor[gray]{0.8} & \cellcolor[gray]{0.8} & B: no dyn. @ $F> -22$ Hz  \\
 & \cellcolor[gray]{0.8} & \cellcolor[gray]{0.8} & C: $F_1^c=17$ Hz\\
 & \cellcolor[gray]{0.8} & \cellcolor[gray]{0.8} & D: no dyn. @ $F_2 < 25$ Hz  \\
 \hline
 &              \;\;\;\;\;A: $F^c=17$ Hz      & \cellcolor[gray]{0.6} \;\;\;\;\;A: $F^c=17$ Hz                   &\;\;\;\;\;\;A: $F^c=12$ Hz\\
Iron/Iron & \;\;\;\;\;B: $F^c=-18$ Hz     & \cellcolor[gray]{0.6} \;\;\;\;\;B: no dyn. @ $F> -25 Hz$   &\cellcolor[gray]{0.8}  \;\;\;\;\;B: no data\\
 &              C/D: $F_{1,2}^c=16$ Hz  & \cellcolor[gray]{0.6} C/D: $F_{1,2}^c=16$ Hz                & \cellcolor[gray]{0.8} C/D: no data\\
 \hline
\end{tabular} \end{spacing}
\caption{Summary of the dynamo/non dynamo regimes observed in the different VKS configurations. Label ``A" corresponds to the ``unscooping" exact counter-rotating case ($F_1=F_2>0$), label ``B" to the ``scooping" exact counter-rotating case ($F_1=F_2<0$), labels ``C" and ``D" to the cases with a single impeller rotating, respectively $F_1>0$, $F_2=0$ and $F_2>0$, $F_1=0$, these two cases being different when two impellers of different materials are used (middle line in the table). Light grey cells are cases where no data is available yet. The values indicated are either the observed critical impellers rotation rate when dynamo is present, or the maximal rotation rate achieved, due to power limitations, without observing dynamo. The correspondance between rotation rate and ${R_m}$, as defined in this paper, is indicated for each configuration. The dark grey cell corresponds to the configuration presented in this paper.}\label{tab:config}
\end{center}
\end{table*}
 
\subsection{Configurations and dynamo capacity}\label{config}

The first observation of the dynamo effect in our experiment was made with inner copper wall, inner ring and iron impellers in place. By removing the inner ring, dynamo action is observed for $F_1,F_2 > 0$, but the main change is that no dynamo regime is observed when both impellers still counter-rotate, but in the opposite (``scooping") direction ($F_1=F_2 < 0$) up to the maximum operational power. The disappearance of the dynamo regime with ``scooping" blades is also observed in kinematic simulations using measured mean velocity fields in an equivalent water experiment~\cite{Pinter:EPJB:09}, although the context is different since the dynamo modes in this case are non-axisymmetric. Note that changing the direction of rotation of the impellers changes the ratio of poloidal over toroidal components of the velocity field because of the curvature of the blades. In the case where the median copper ring is present, dynamo regimes are observed for both directions of rotation of the impellers, although the thresholds differ by about $6\%$. These observations indicate that {some aspects of }the flow structure (poloidal/toroidal ratio, position of recirculation loops, stability of the shear layer, level of fluctuations, ...) do play a role in the onset of the instability. 

For the configuration with the inner wall removed, the dynamo threshold is reached for a rotation rate 1.4 times lower for exact counter-rotation, compared to the case with inner wall, corresponding to the same critical ${R_m}$, as defined in section~\ref{setup}. 

Other experimental configurations have been studied, in which the iron impellers are replaced by stainless steel (SS) ones (with and without inner walls and ring). In these cases, no dynamo is observed at the highest $R_m$ achievable with our experiment. A hybrid configuration has also been tested in which one of the impellers is made of soft iron and the other of stainless steel (no inner walls or ring). In the exact counter-rotating case, dynamo generation is observed above $F^c=15$ Hz. When the iron impeller only is rotating and the stainless steel impeller is kept at rest, dynamo is observed above a different critical rate $F^c=17$ Hz and no dynamical regimes are evidenced in this case. When only the stainless steel impeller is rotating, no dynamo could be observed up to a maximum rotation rate of 25 Hz. In addition, no dynamo is observed in exact corotation.

Table~\ref{tab:config} presents a summary of these results, including the dynamo thresholds based on the rotation frequency, obtained in the various configurations which produced a dynamo. {The study presented in this paper corresponds to one specific configuration, where the parameter space has been explored in details. It includes iron impellers and the inner copper cylinder, without the copper ring in the mid-plane. Nevertheless, similar parameter spaces have been observed in all VKS flows driven by 2 iron impellers.} For instance, when the inner cylinder is removed, the parameter space still has non-dynamo to dynamo transitions with supercritical bifurcations to statistically stationary magnetic fields and regimes with oscillations and reversal are observed, with the restriction that only one window of time dependent behavior is evidenced.

\section{Parameter space and bifurcations}
%%%%%%%%%%%

\begin{figure}
\centerline{\includegraphics[width=8.5cm]{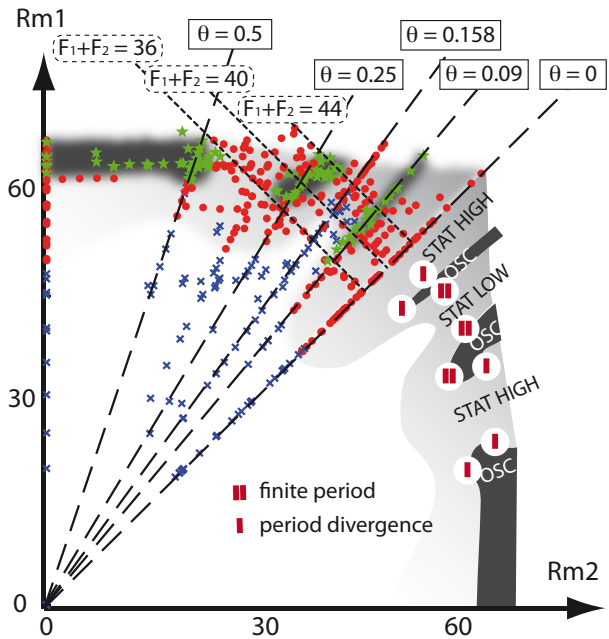}}
\caption{$({R_m}_1,{R_m}_2)$ parameter space explored, varying independently the rotation rate of each impeller (top left, where blue crosses indicate no dynamo regimes, red circles stationary dynamo regimes and green stars time dependent regimes), and schematic of accessible dynamical regimes (bottom right) -- boundaries between different regimes are voluntarily smoothed due to lack of experimental resolution. The type of transition is also indicated, as well as some paths along $F$ or $\theta$ constant, followed during the experimental investigations. }
\label{fig1bis}
\end{figure}

The flow and dynamo configurations spanned in the results reported here are summarized in Fig.\ref{fig1bis}. It shows the $({R_m}_1,{R_m}_2)$ parameter space (only in the case $F_1,F_2>0$), where each ${R_m}$ is defined based on the rotation rate of one impeller. {Since we observe almost no asymmetry in this parameter space under the exchange $F_1 \leftrightarrow F_2$, the actual experimental points are represented only in the upper left half of the plane (see Fig.~\ref{fig5}b for an estimate of the experimental imperfections).} Color codes are used, corresponding to the dynamo regime observed at each location, broadly characterized as non dynamo, (statistically) stationary magnetic field and time-dependent regimes. We also indicate the paths that have been followed in the experimental measurements at $F$ or $\theta$ constant (note that $\theta$=constant also corresponds to $F_1/F_2=$constant) along which dynamo transitions are discussed in more details below. In the lower right part of the graph in Fig.~\ref{fig1bis}, a schematic view of the regimes is sketched.

\begin{figure}
\centerline{\includegraphics[width=8.5cm]{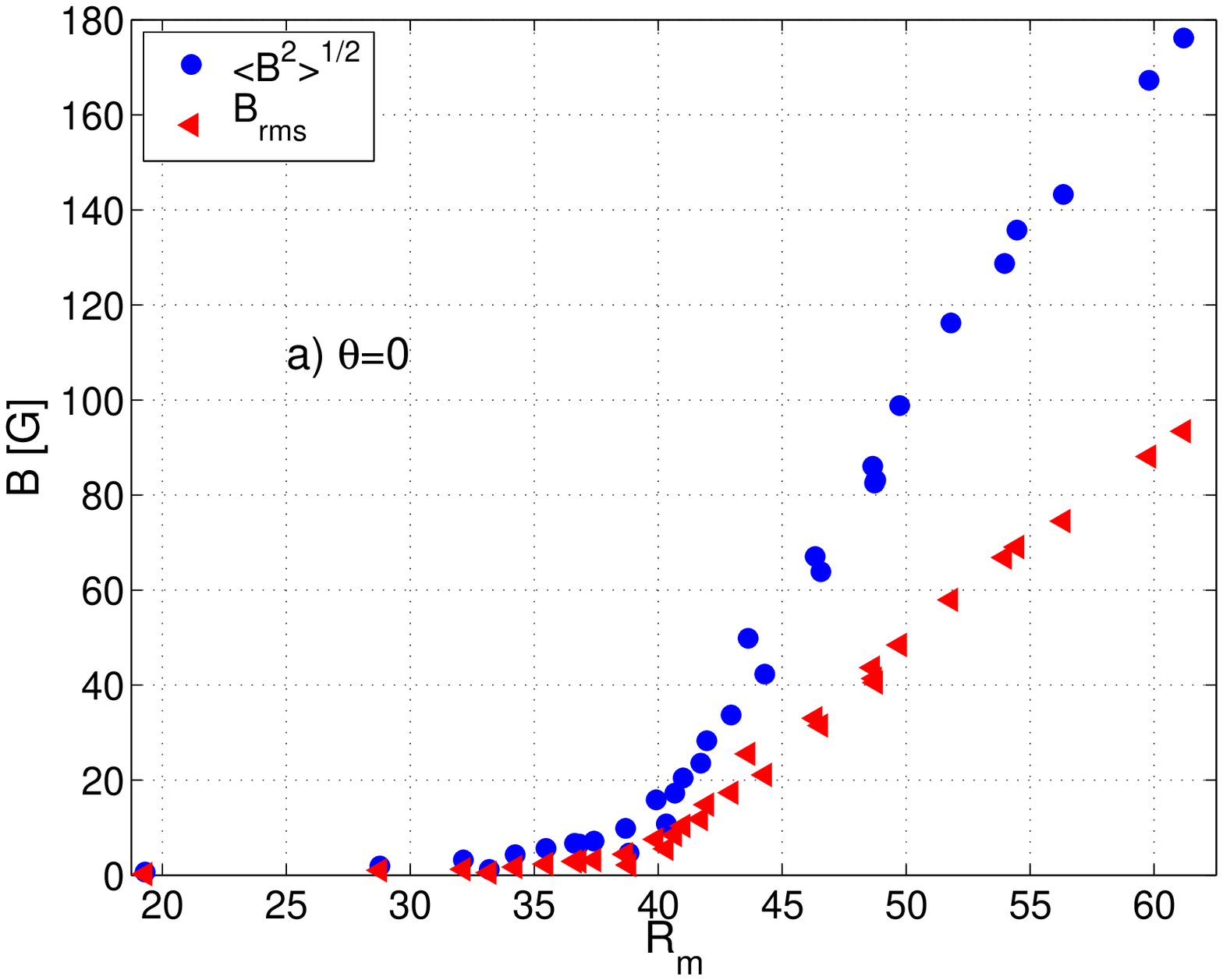}}
\centerline{\includegraphics[width=8.5cm]{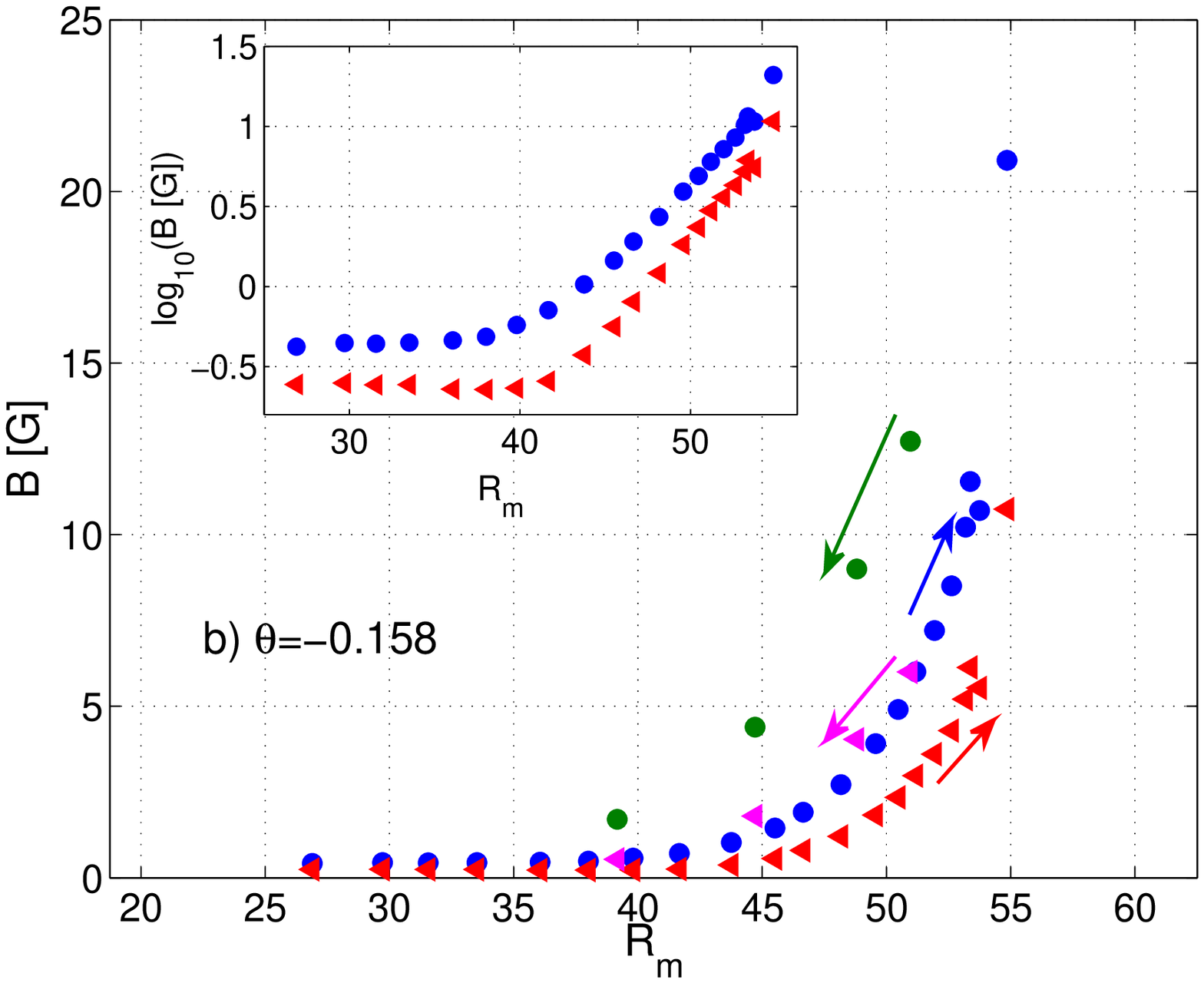}}
\caption{Bifurcation to stationary dynamos, magnetic field amplitude and standard deviation. a) Exact counter-rotation, $\theta=0$, STAT HIGH regime; b) $\theta=-0.158$, STAT LOW regime. {In this case, as indicated by the arrows, data points measured with increasing $R_m$ are shown in blue (amplitude) and red (standard deviation), while data points measured with decreasing $R_m$ are shown in green (amplitude) and magenta (standard deviation).} The inset shows the same plot with vertical log scale, in the case of increasing $R_m$.}
\label{fig2}
\end{figure}

We start by exploring the bifurcations that develop at a constant asymmetry parameter $\theta$ as the magnetic Reynolds number is increased. At all investigated values of $\theta$, we observe that increasing ${R_m}$, the first instability is a supercritical bifurcation to a stationary dynamo, with time-dependent regimes possibly developing at larger $R_m$ values. {We note that the possibility of a direct bifurcation to time-dependent dynamo cannot be excluded, for instance near $\vert\theta\vert\sim 0.12$ where non dynamo and oscillating regimes are observed in Fig.~\ref{fig1bis} in very close proximity.} We first discuss the case where time dependent regimes do not appear, the dynamo field remaining statistically stationary at the highest $R_m$ values achieved in the experiment (due to power/torque limitations). Figure~\ref{fig2} shows examples of this type. The amplitude and standard deviation of the magnetic field are plotted as a function of ${R_m}=({R_m}_1+{R_m}_2)/2$. In the case of exact counter-rotation ($\theta=0$, top figure), a clear transition to dynamo can be observed at ${R_m^c}\sim 39$ after an initial slowly growing phase (${R_m} < 38$), which we interpret as induction from the ambient magnetic field. Above the threshold, the amplitude of the magnetic field is observed to grow linearly with $R_m-{R_m}_c$ {up to $R_m\sim 55$, after which the trend is less clear}. This is different from the case when the inner ring is present~\cite{Monchaux:PoF:09}, where a best fit of dynamo field growth lead to a power law increase $(R_m-{R_m}_c)^\alpha$, with $\alpha \sim 0.77$. {Note that, due to the imperfection of the bifurcation curve already discussed in presence of the inner ring, the determination of the exponent depends strongly on the chosen value for the threshold $R_m^c$, leading to a larger uncertainty}. Simple arguments from bifurcation theory~\cite{Strogatz:94} would lead one to expect an $\alpha=1/2$ value. For such a bifurcation from a turbulent state, the existence and universality of critical exponents is still an open problem.

At $\theta=-0.158$ (bottom figure), we also observe a supercritical bifurcation, but the amplitude of the dynamo field is much lower, by almost an order of magnitude. We thus call this regime ``STAT LOW" as opposed to the ``STAT HIGH" regime observed in exact counter-rotation. The change in behavior from an induction regime to a low field dynamo regime is best seen in the lin/log plot in the inset, showing a clear break in the slopes at about ${R_m}\sim 41-42$. {It is a low amplitude dynamo, and the effect of the slight magnetization of the iron disks cannot be ignored; hence the magnetic field is different if the measurement is made after a dynamo run (green and magenta symbols in~\ref{fig2}b -- measured when decreasing $R_m$) or following a non dynamo configuration (blue and red symbols-- measured when increasing $R_m$).} Note that induction effects from the ambient field are also present, which can add up to the starting dynamo. Because in this case the dynamo is a low amplitude one, both effects can be of the same order of magnitude, so that the behavior near threshold is non trivial. The standard deviations follow the same evolution, with an amplitude about half of that of the mean fields.

\begin{figure*}
{\begin{minipage}{0.99\textwidth}
\centering
\includegraphics[width=.32\textwidth]{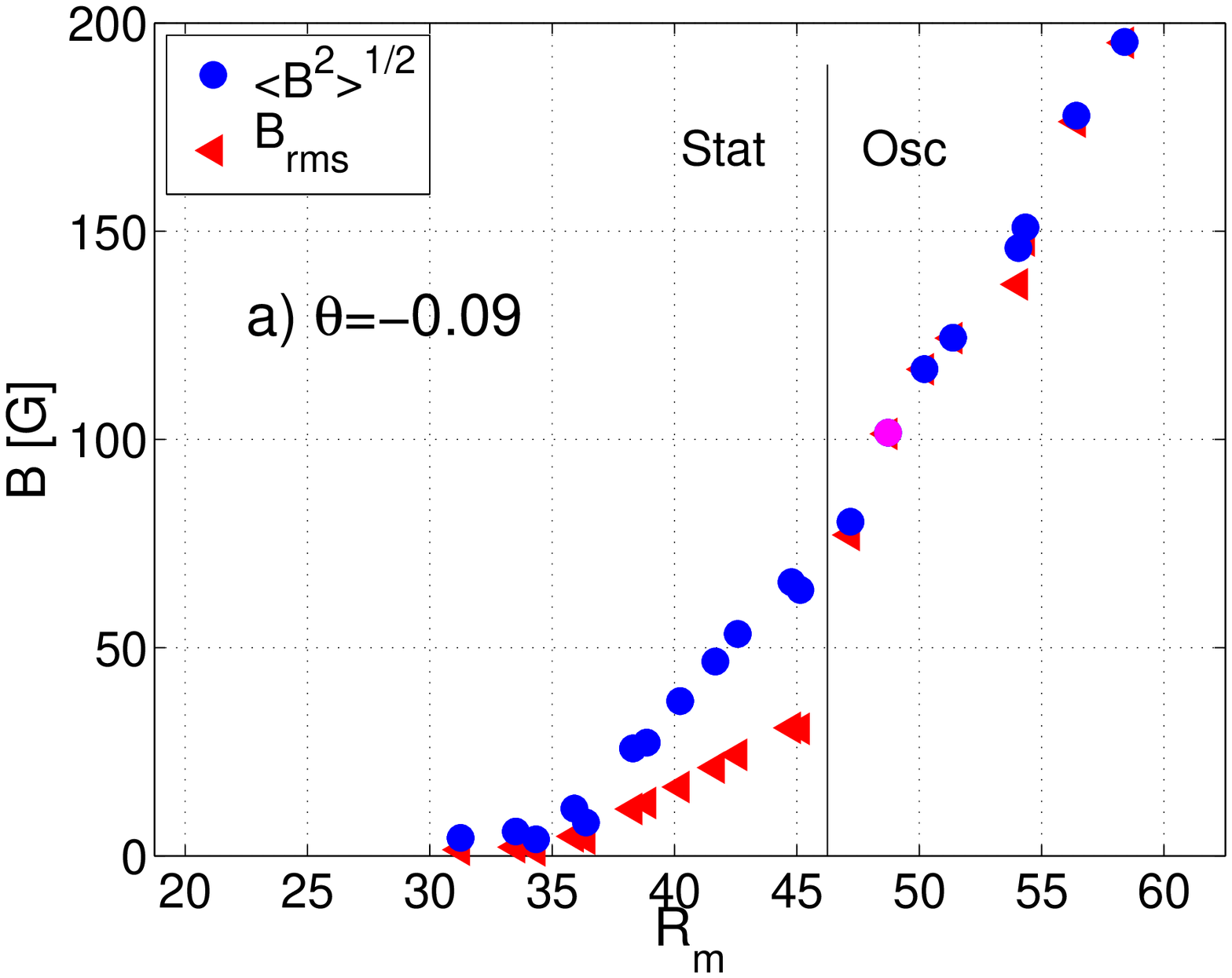}
\includegraphics[width=.32\textwidth]{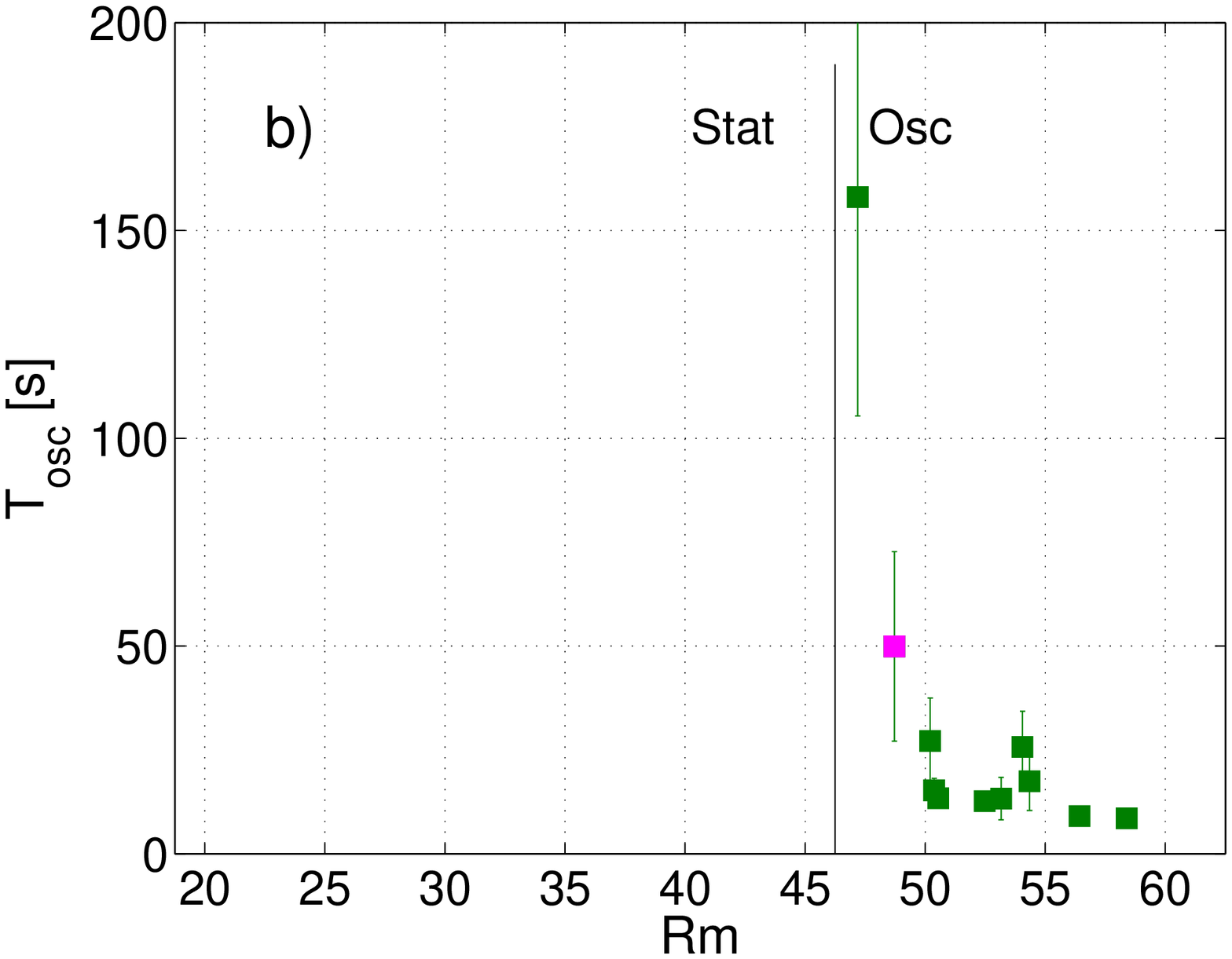}
\includegraphics[width=.32\textwidth]{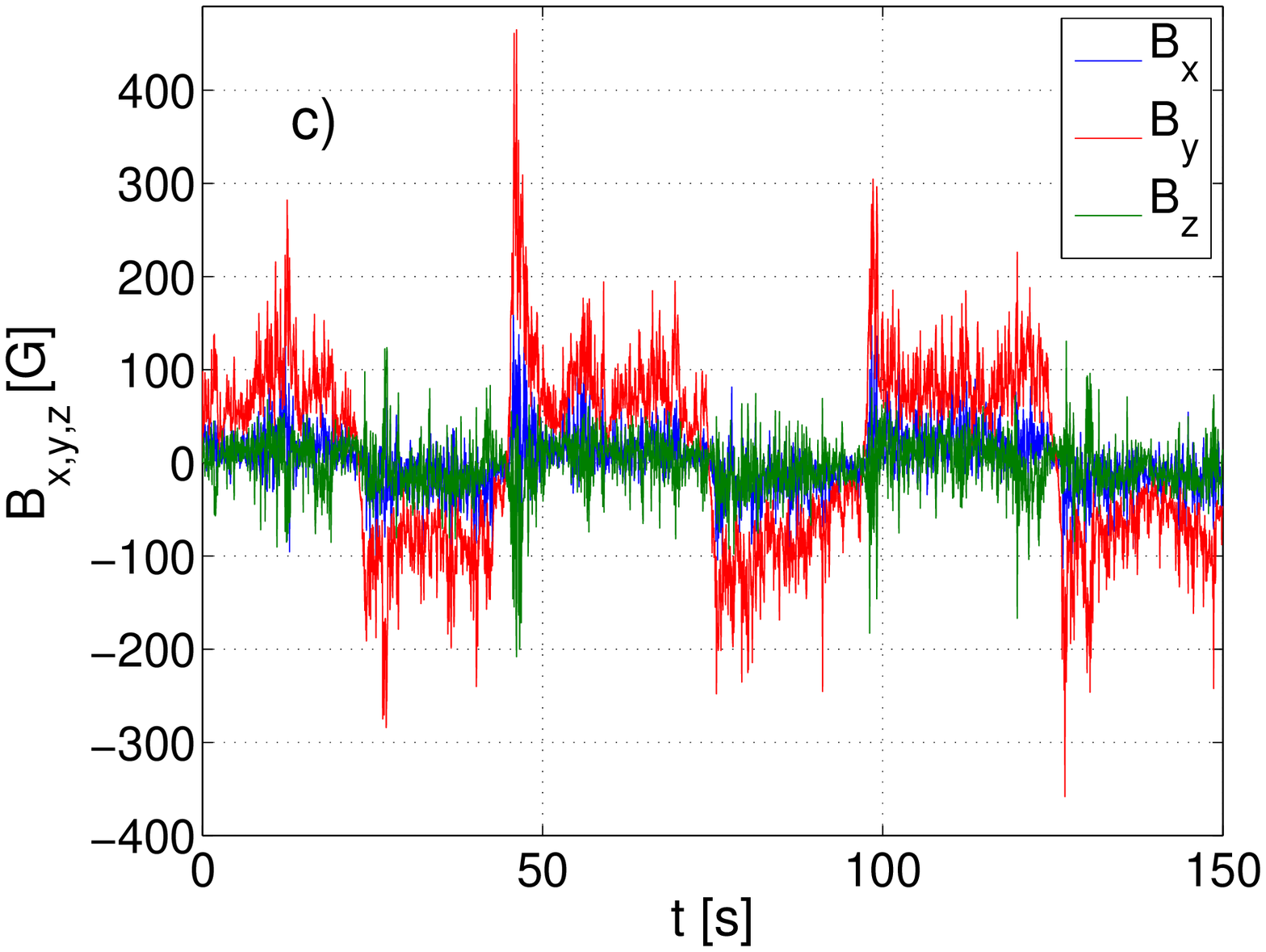}\\
\includegraphics[width=.32\textwidth]{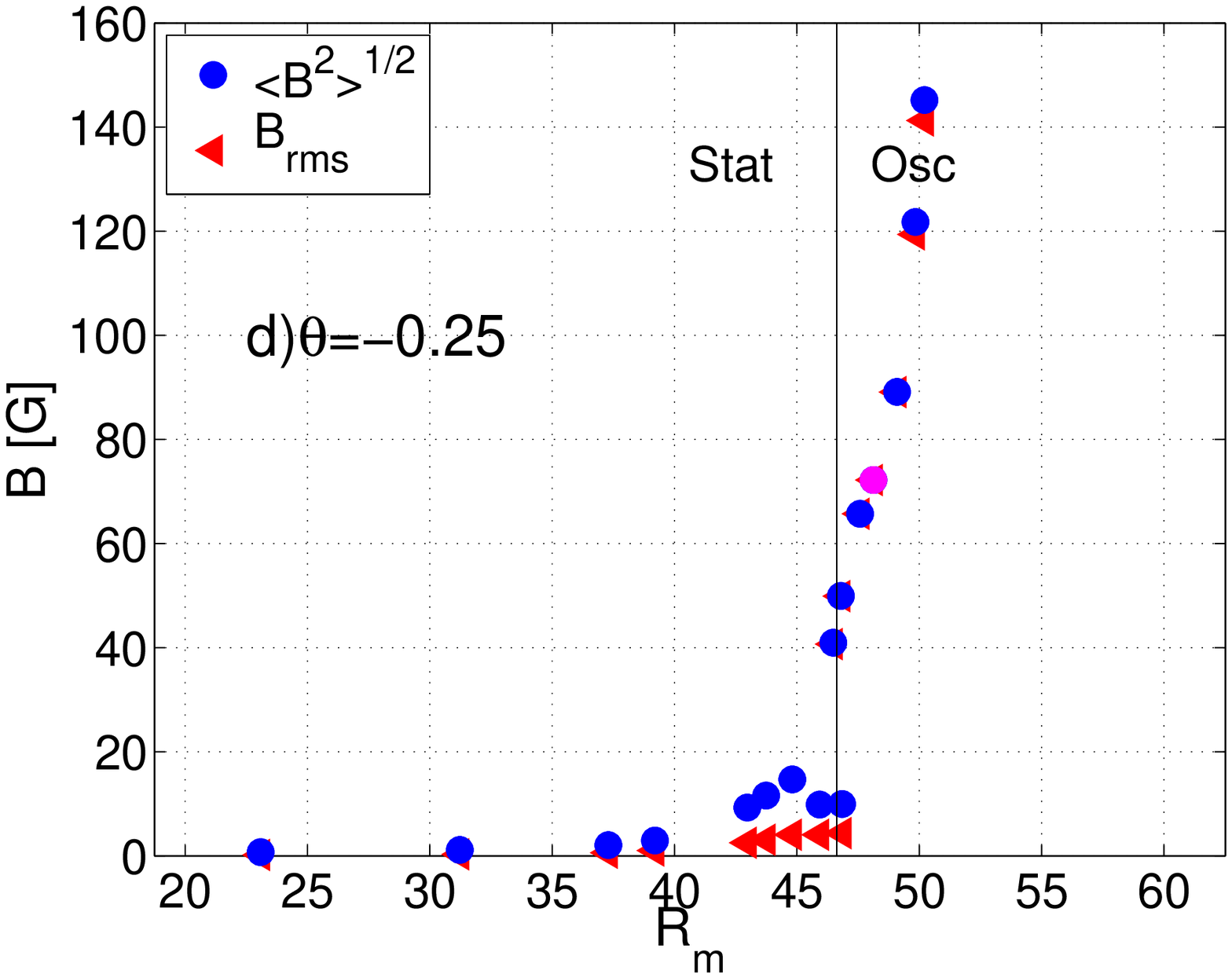}
\includegraphics[width=.32\textwidth]{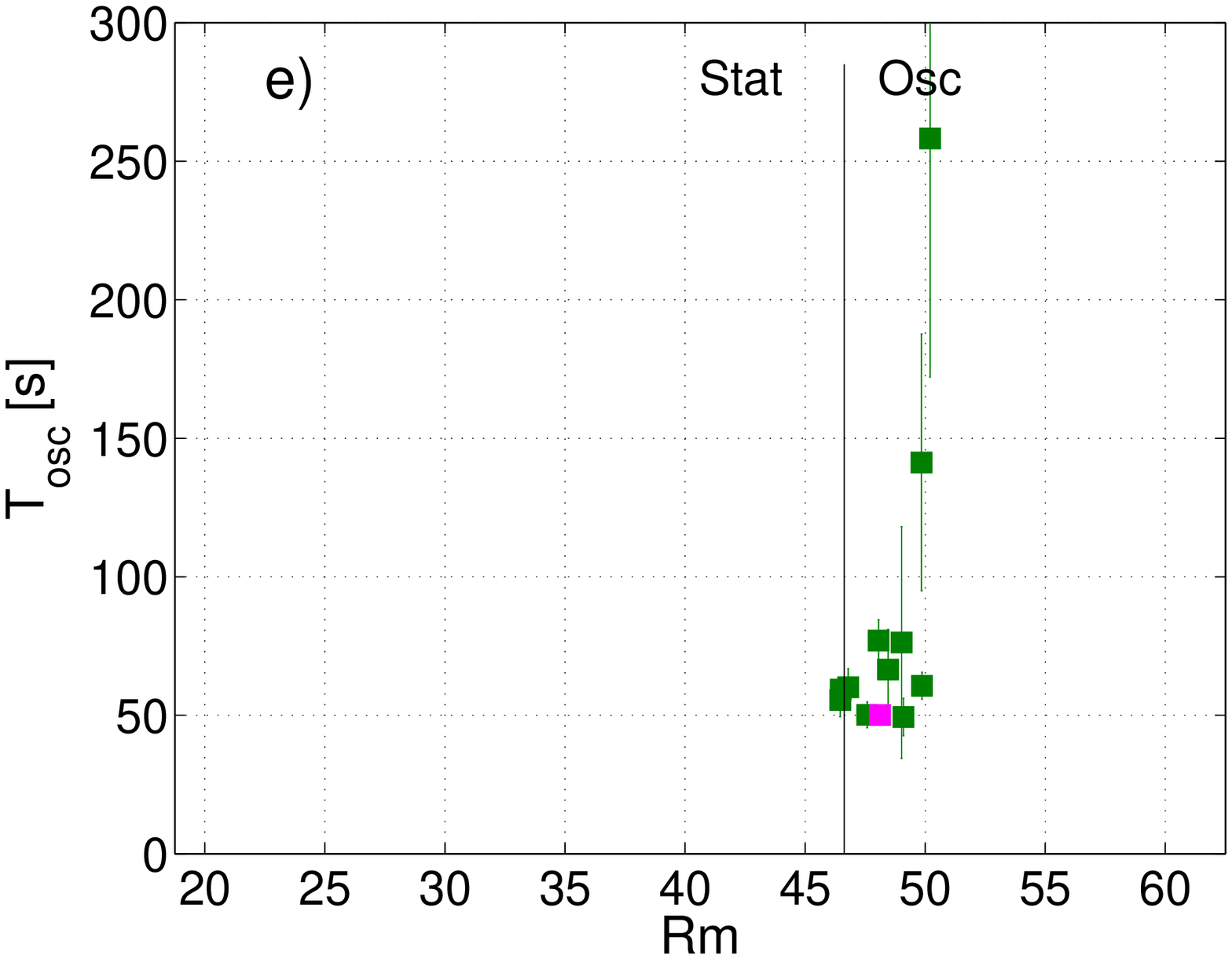}
\includegraphics[width=.32\textwidth]{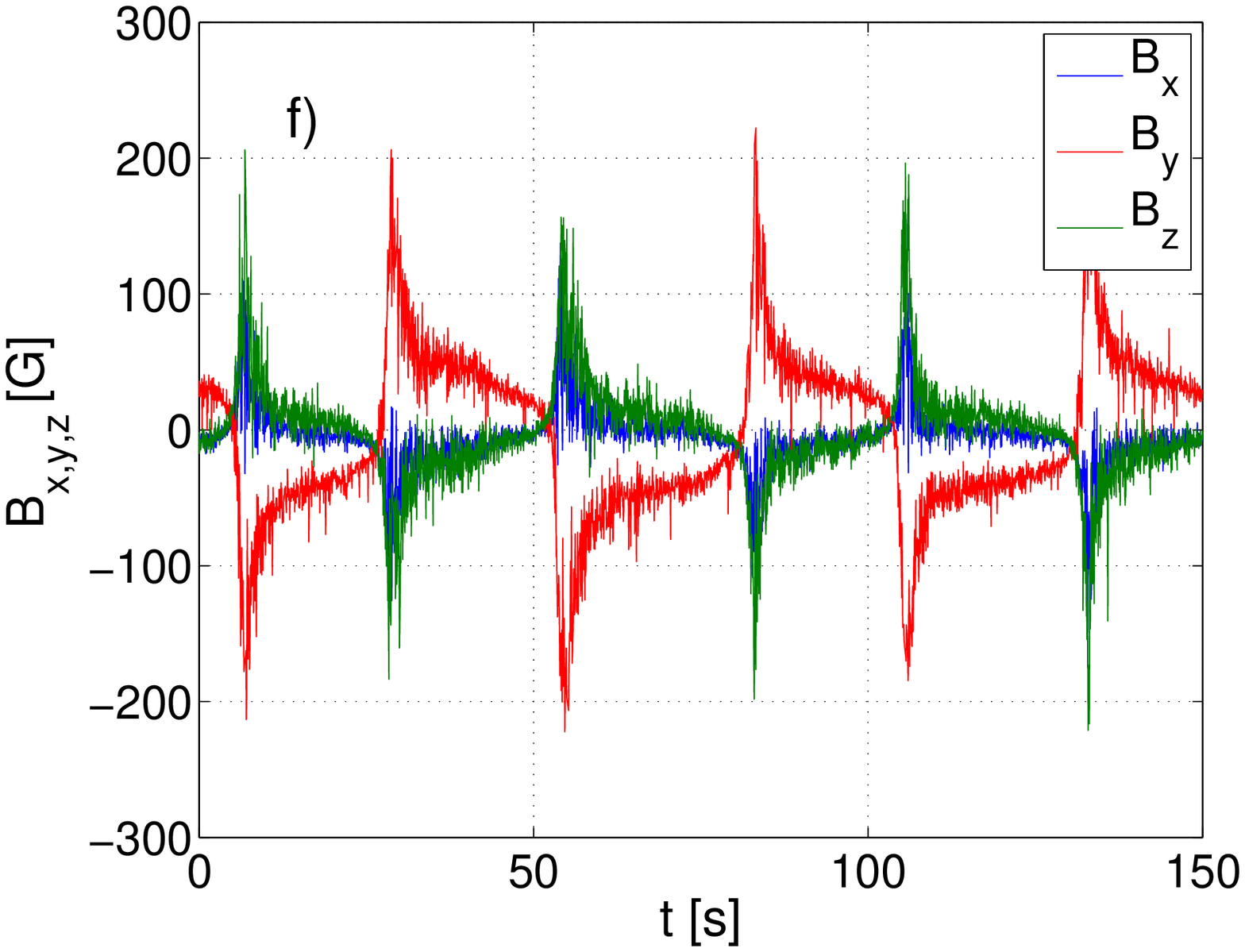}\\
\includegraphics[width=.32\textwidth]{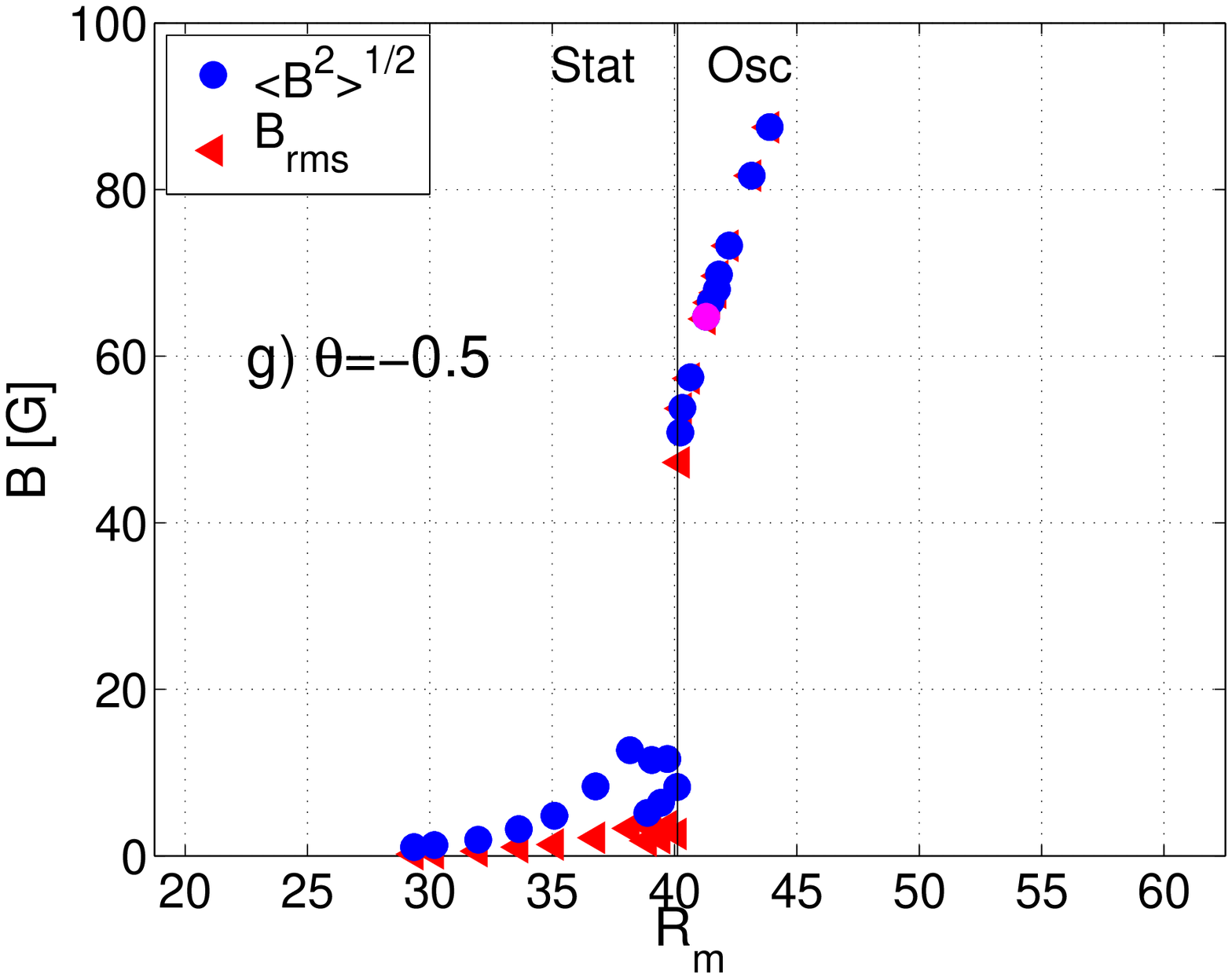}
\includegraphics[width=.32\textwidth]{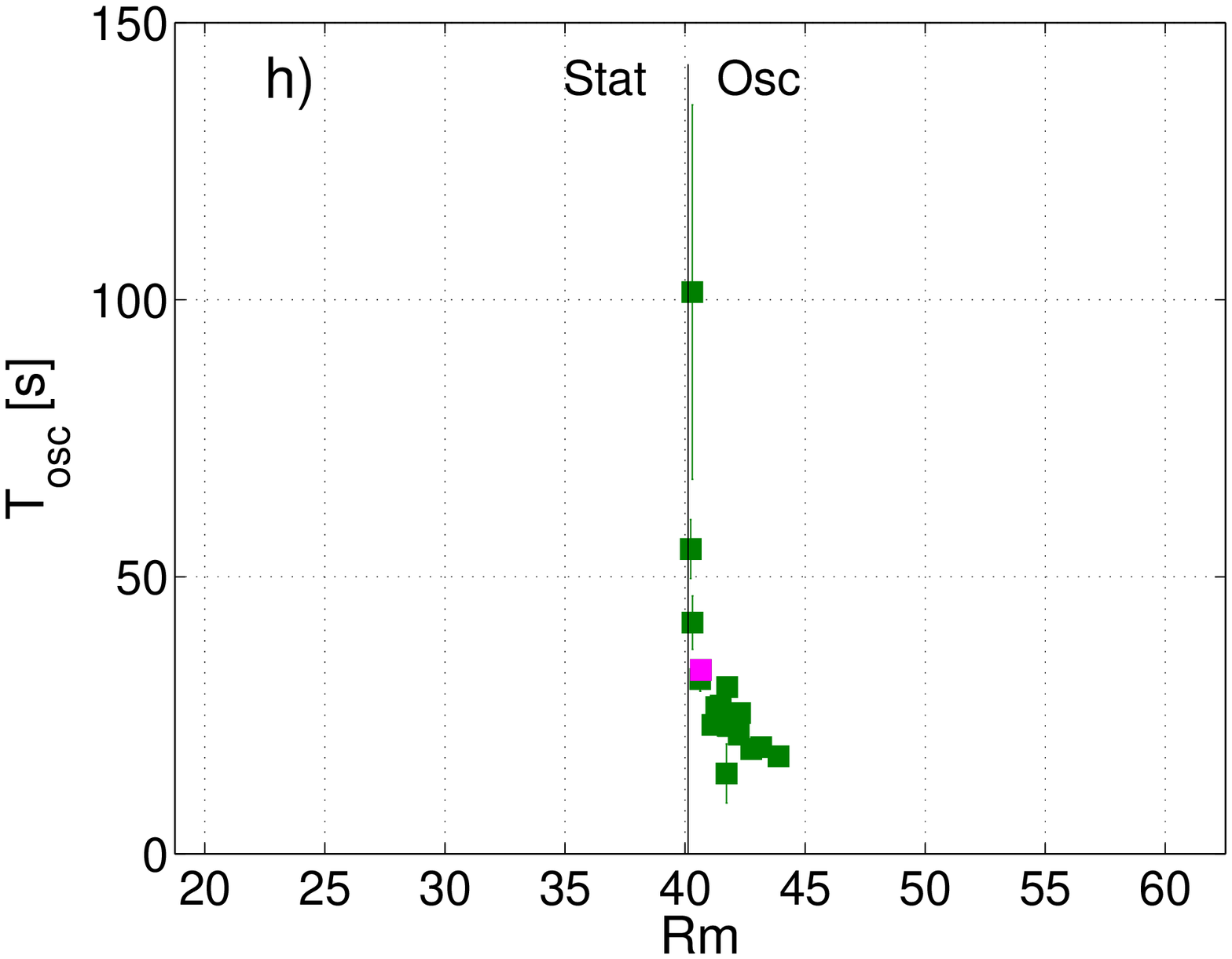}
\includegraphics[width=.32\textwidth]{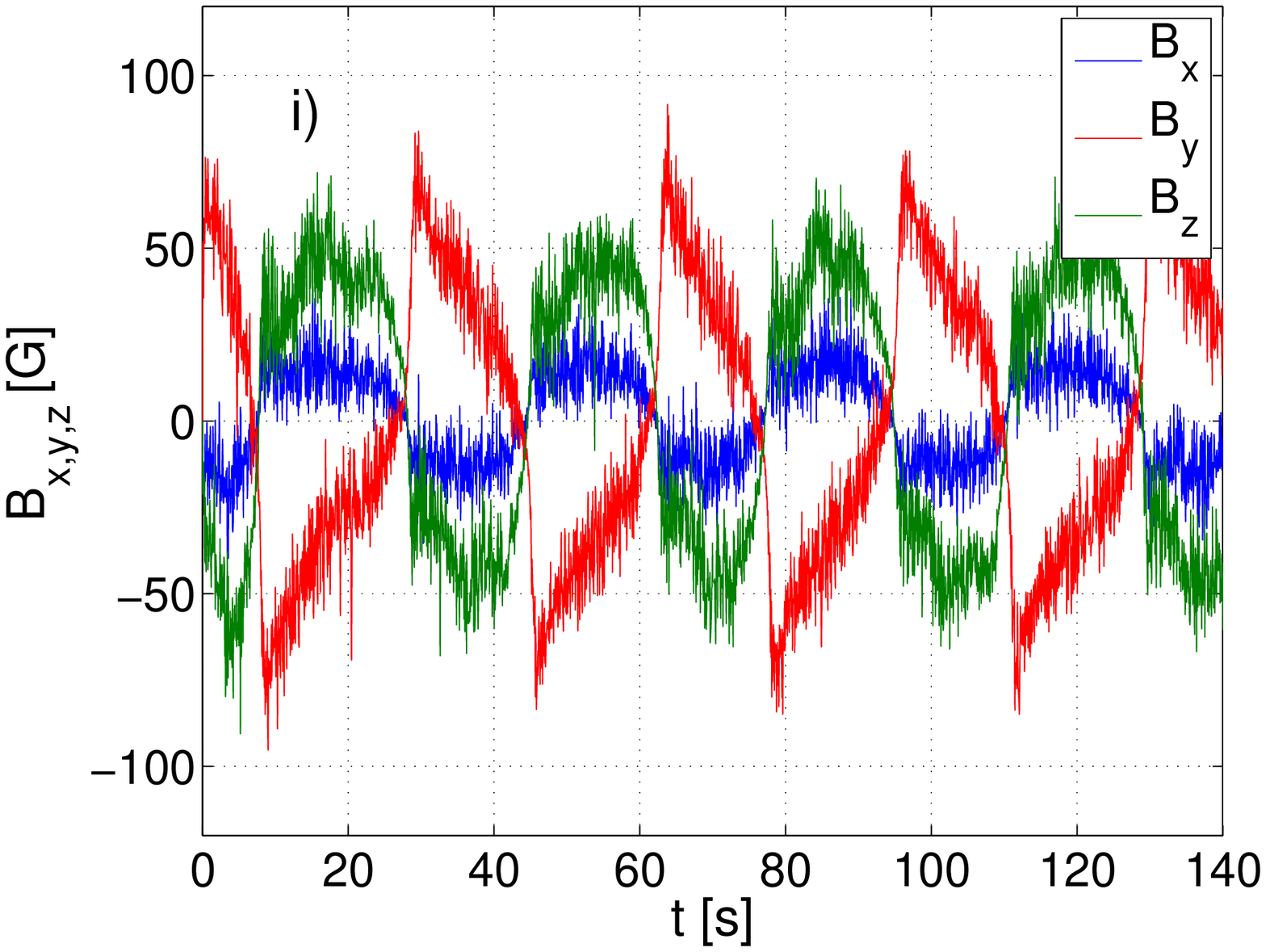}\\
\caption{Examples of dynamo bifurcations with a secondary bifurcation to a time-dependent regime (left) and corresponding periods of oscillation (middle). The error bars represent the standard deviation for the series of measurements. In the case where only one (or one half) period is observed, the error has been arbitrarily set to one third of the period. The separation between stationary and oscillatory regimes is indicated by a vertical line. For each case, one point (magenta symbol in each of the left and middle plots) has been chosen in the oscillating region, for which the corresponding time signal is shown in the right plots.}
\label{fig3}
\end{minipage}}
\end{figure*}

Fig.~\ref{fig3}a,d,g shows bifurcation curves for other values of $\theta$, for which we observe the development of time dependent regimes as $R_m$ is increased. A stationary dynamo is first generated at ${R_m}_{c1} \sim 25-30$, then oscillations develop as ${R_m} > {R_m}_{c2} \sim 30-40$ (examples of the time signals shown in Fig. ~\ref{fig3}c,f,i). For the larger values of $\theta$, the dynamical change is also associated with a discontinuity in the amplitude of the magnetic field. It indicates that this second bifurcation is not supercritical, although the observation of the discontinuity may depend on the choice of the order parameters. When the regime becomes oscillatory, the mean value vanishes for each component, as can be seen by the equality between the field amplitude and its standard deviation. Another interesting observation, evidenced in Fig.~\ref{fig3}b,e,h, is that at low ($\theta=-0.09$) and high ($\theta=-0.5$) asymmetry, the period diverges at the secondary bifurcation threshold. At the intermediate value $\theta=-0.25$ the oscillations develop with a finite period.

\section{Transitions between regimes}
%%%%%%%%%%%

\begin{figure}
{\begin{minipage}{0.5\textwidth}
\centering
\includegraphics[width=.98\textwidth]{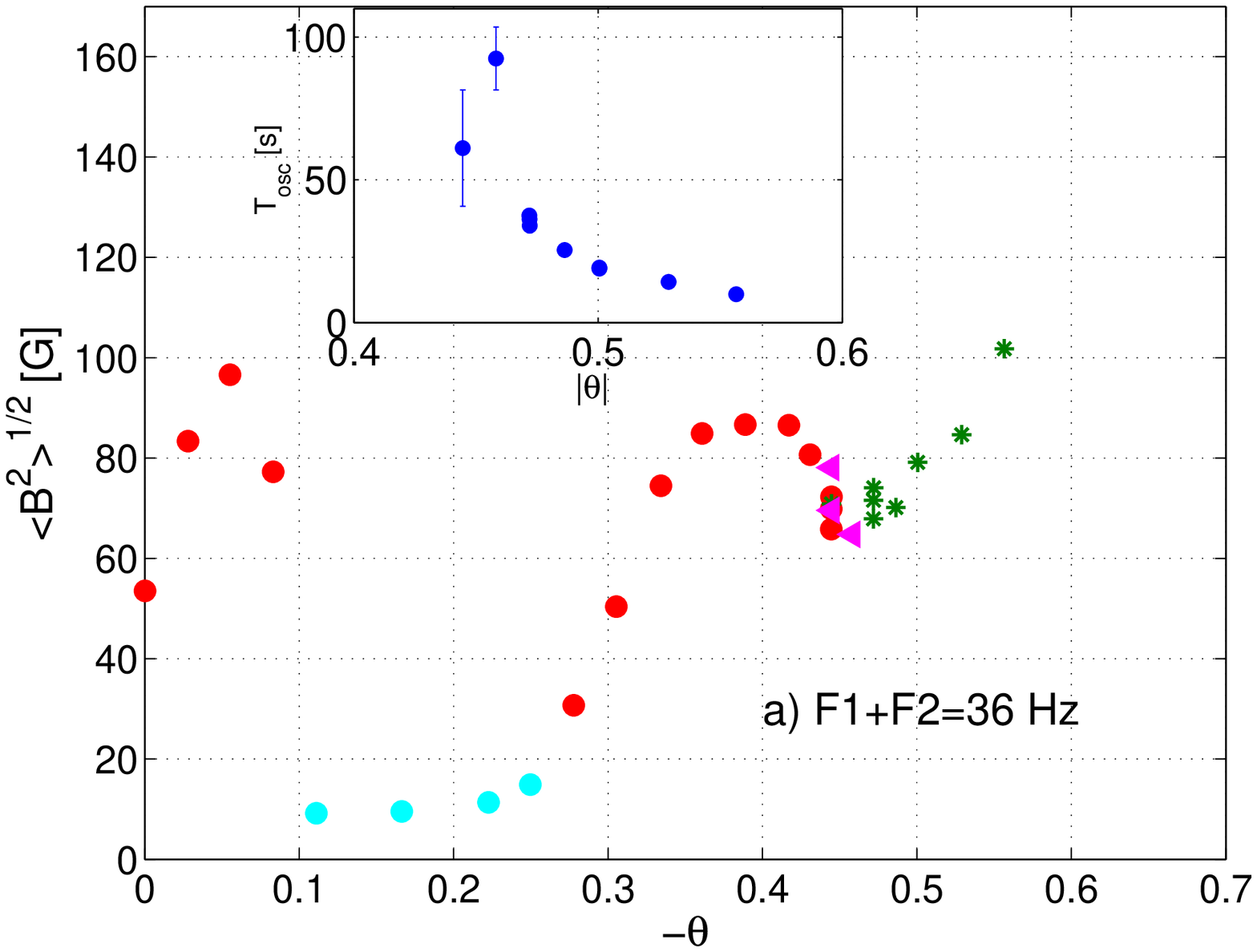}\\
\includegraphics[width=.98\textwidth]{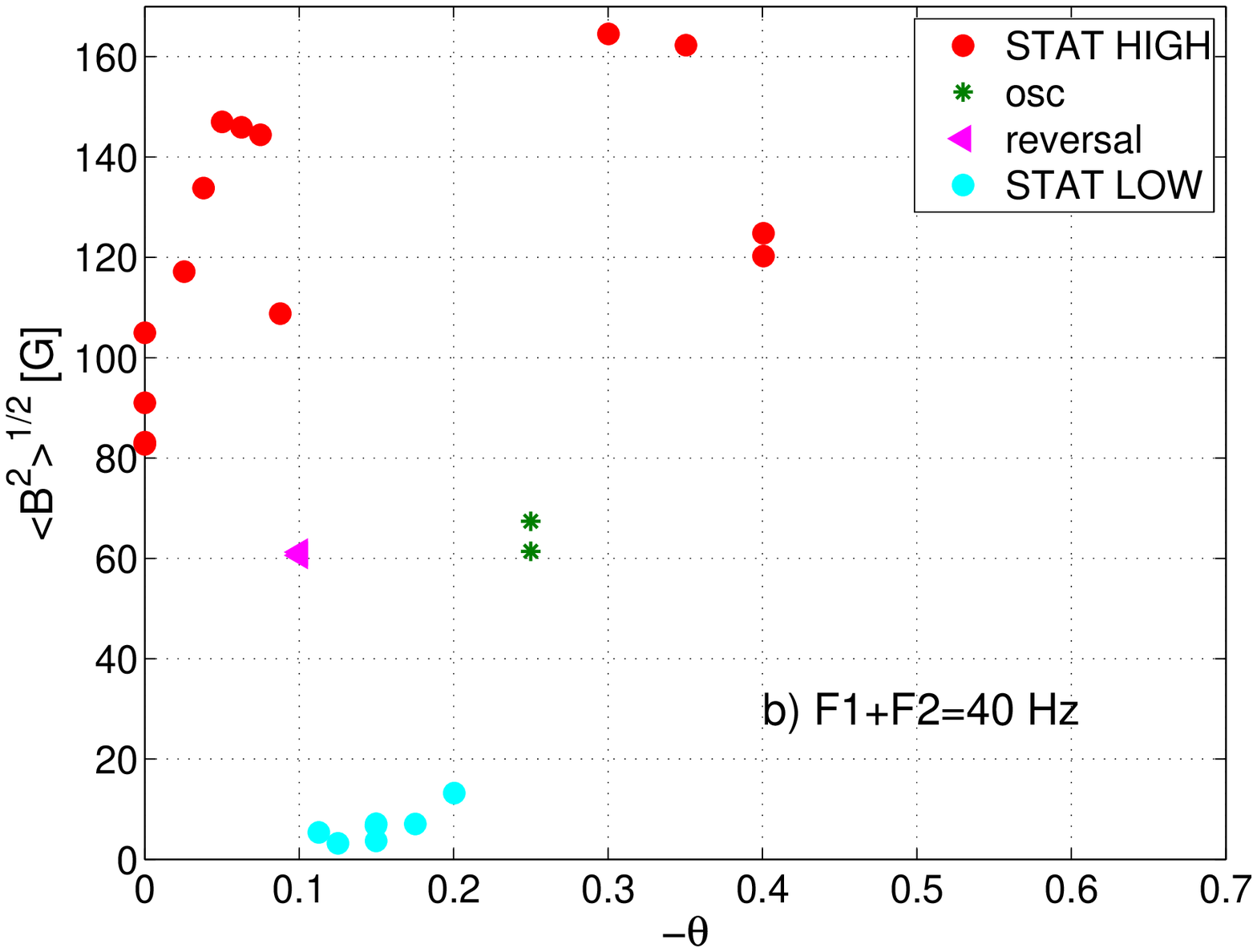}\\
\includegraphics[width=.98\textwidth]{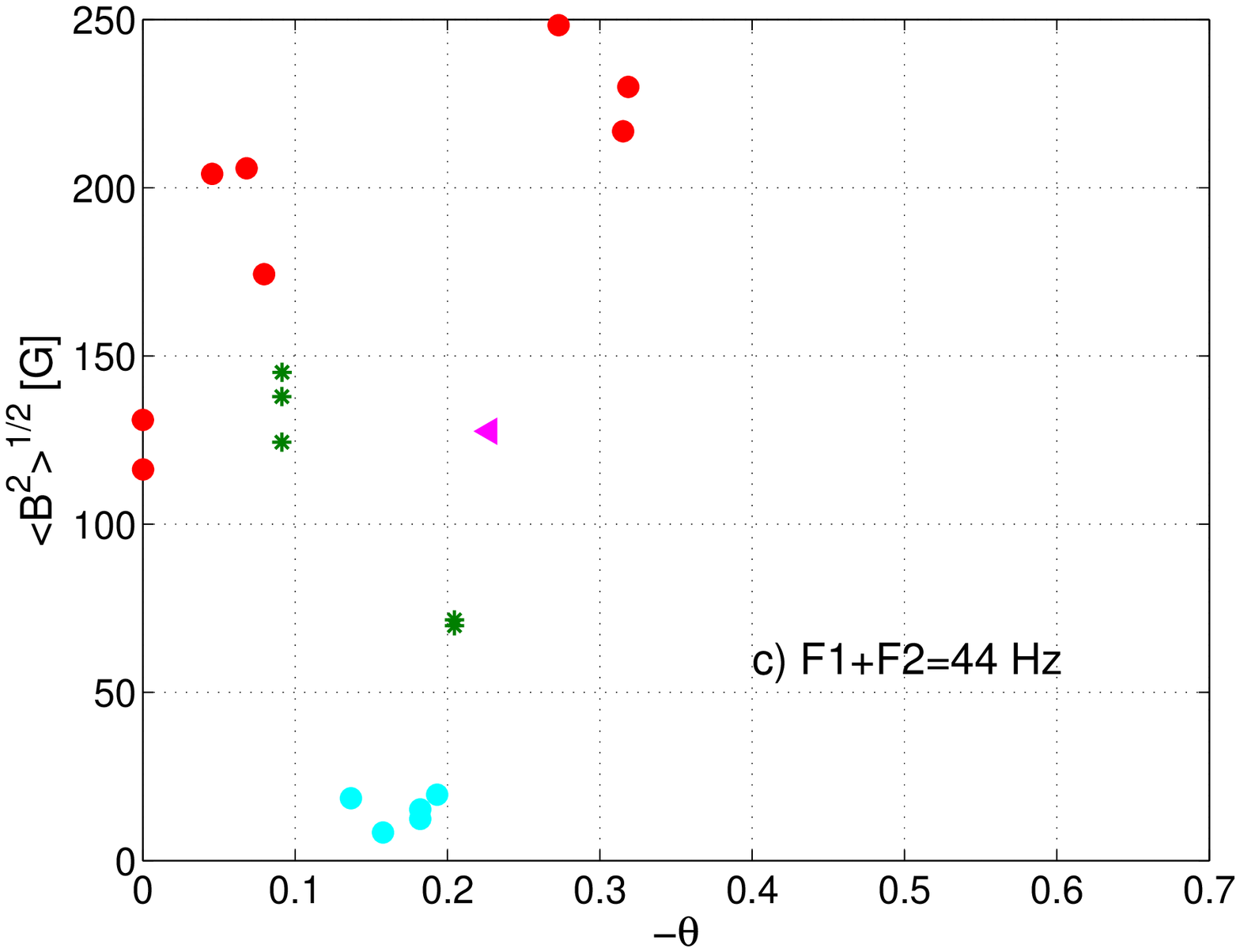}
\caption{(a) Transition plots as asymmetry ($\theta$) is varied at fixed $F$. a) $F_1+F_2=36$~Hz. Color code as in (b). The inset shows the corresponding evolution of the period of oscillation in the oscillating/reversing regimes. (b) $F_1+F_2=40$~Hz. (c) $F_1+F_2=44$~Hz. Color code as in (b).}
\label{fig4}
\end{minipage}}
\end{figure}

\begin{figure}[t!]
\centerline{\includegraphics[width=9.2cm]{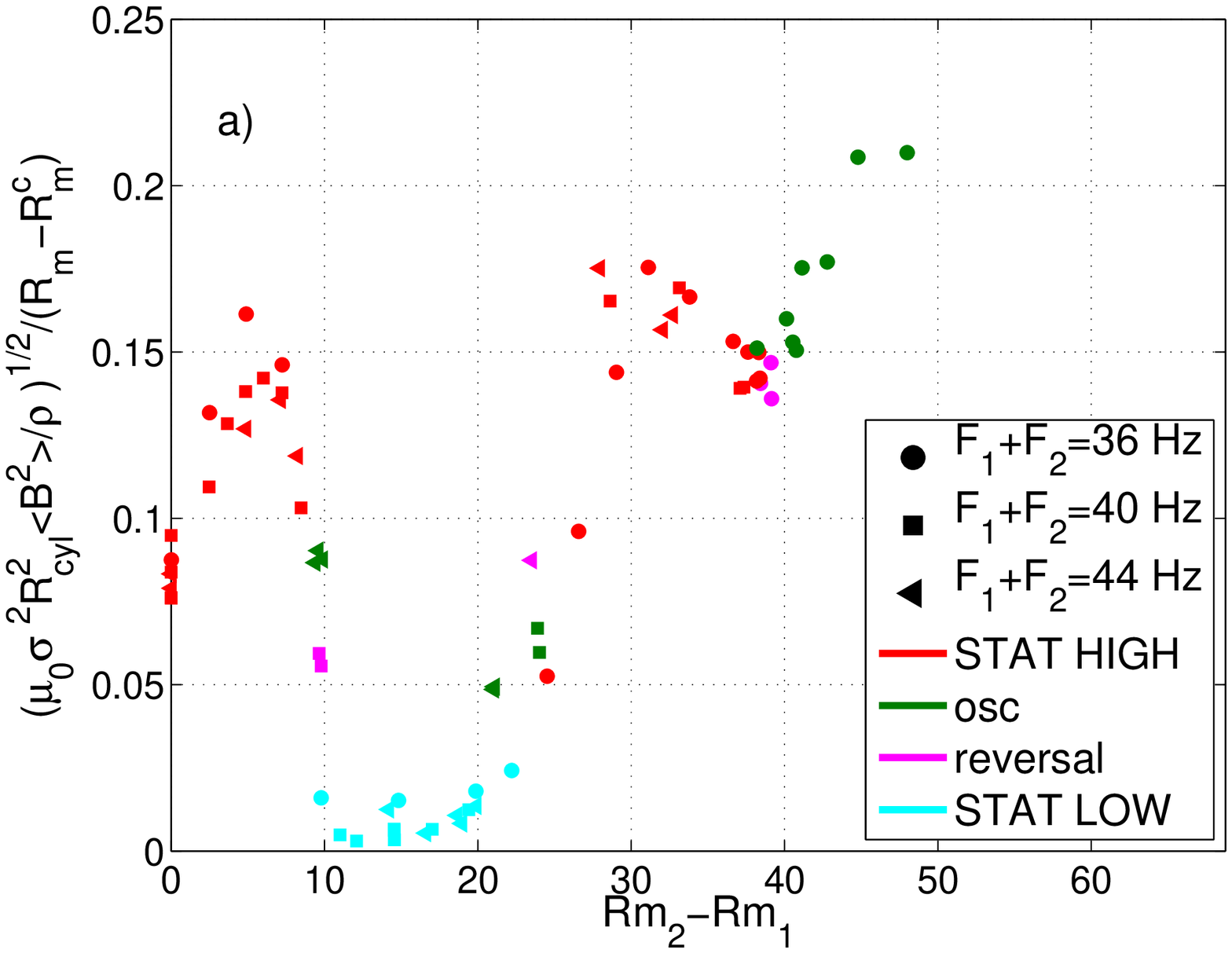}}
\centerline{\includegraphics[width=9.2cm]{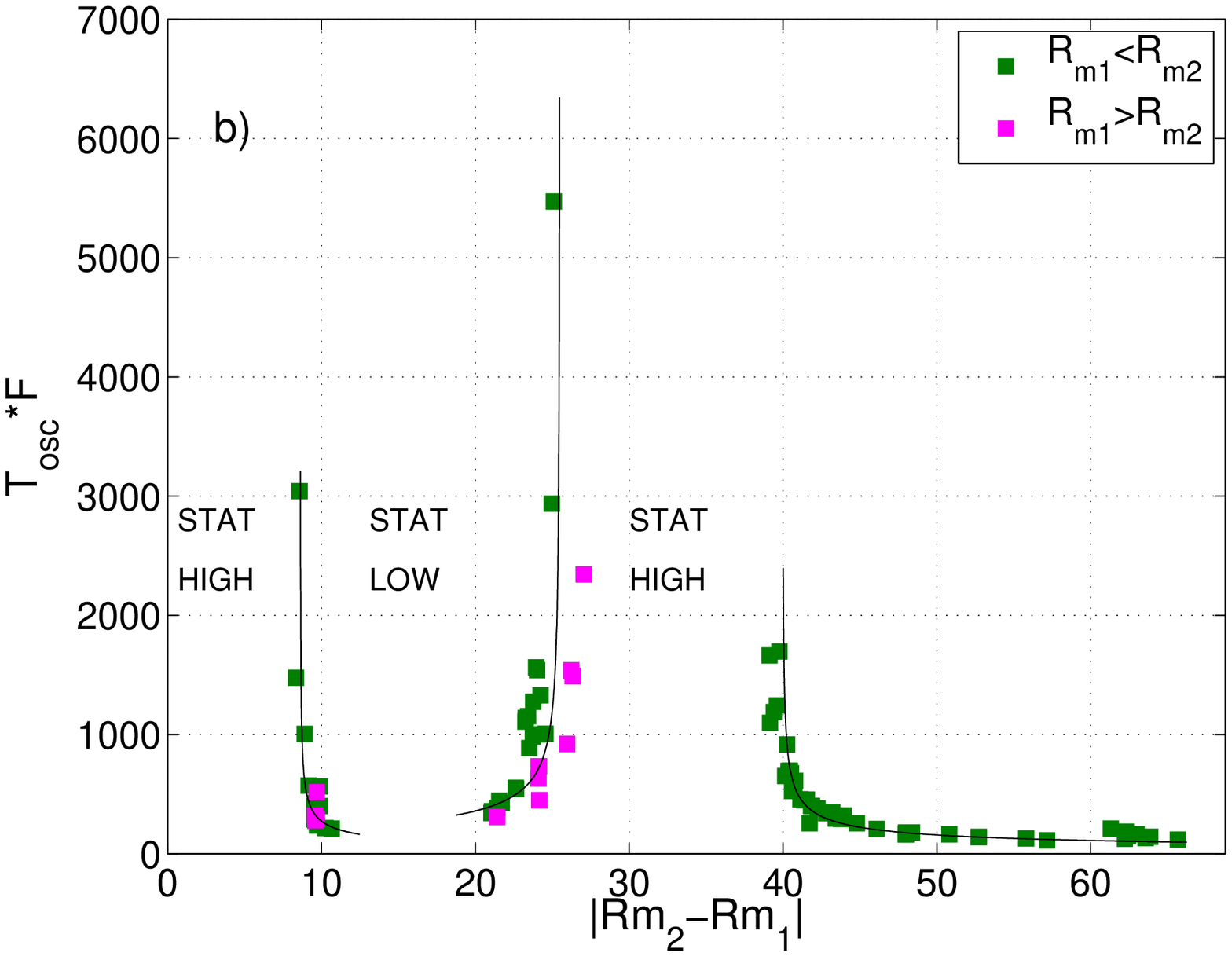}}
\caption{a) Rescaling of the 3 transition plots of figure \ref{fig4}, with $\Delta {R_m}={R_m}_2-{R_m}_1$ for the abscissa and $\vert B\vert\sigma R_{\rm cyl}\sqrt{\mu_0/\rho}/(R_m-R_m^c)$ for the ordinate ($R_m^c=39$). Symbols correspond to the different values of $R_m$, colors show the different dynamo regimes. b) Evolution of the period of oscillations on either side of several windows of stationary dynamos.  {In this plot, since the oscillations period are not expected to depend on the sign of $\Delta {R_m}$ all data points are shown, using different colors for points corresponding to $\Delta {R_m}$ positive and negative.} The solid lines are $(\vert\Delta {R_m}\vert-\vert\Delta {R_m}^c\vert)^{-1/2}$ fits, where $\Delta {R_m}^c$ is the value of $\Delta {R_m}$ at the onset of each of the 3 transitions that display a period divergence. }
\label{fig5}
\end{figure}

The observation of a variety of dynamo regimes when the asymmetry is varied can also be explored in the parameter space along lines of constant $F$, at varying $\theta$ values. This corresponds to a fairly constant global magnetic Reynolds number ${R_m}$, save for small changes due to variations in the temperature of the sodium. We first consider the behavior for $F_1+F_2=36$ Hz (${R_m}=34.8\pm0.6)$ as $\vert\theta\vert$ increases -- shown in Fig.\ref{fig4}a. Close to exact counter-rotation ($\theta\approx 0$), the dynamo is a STAT HIGH type. Then, the region $\theta \in [-0.25,-0.1]$ corresponds to the STAT LOW regime (identified as a STAT LOW dynamo when the field amplitude is larger than 3 G), before reaching again a STAT HIGH regime, as $\theta$ increases. At higher values ($\theta < -0.45$) the stationary dynamo looses its stability for a regime with random reversals before turning into an oscillating one. As shown in the inset, as the oscillating/reversing regime approaches the stationary regime (decreasing $\vert\theta\vert$), the period at the transition diverges. As ${R_m}$ is increased, the main change is the occurrence of reversal and oscillating regimes on each side of the STAT LOW region. This is confirmed in Fig.~\ref{fig4}b and c, where the transition plots are shown respectively in the case $F_1+F_2=40$ Hz (${R_m}=38.6\pm0.5$) and $F_1+F_2=44$ Hz (${R_m}=41.8\pm0.7$). Note that the oscillating region at high $\theta$ observed in Fig.~\ref{fig4}a could also be present at higher forcing, but it was not possible to reach regimes at $\theta<-0.4$ for $F\ge 40$ Hz, due to torque limitations on the motors driving the impellers. It is interesting to observe that these transition plots display a similar shape. Actually, a qualitative collapse of the 3 plots shown in Fig.~\ref{fig4} can be obtained (see Fig.~\ref{fig5}a) if the variable $\Delta {R_m}={R_m}_2-{R_m}_1$ is used for the abscissa and the amplitude of the magnetic field in the ordinate is normalized by $(1/\sigma R_{\rm cyl})\sqrt{\rho/\mu_0}(R_m-R_m^c)$ (with $R_m^c=39$), as suggested from the linear growth observed in Fig.~\ref{fig2}a. Similarly to the $\theta$ variable, $\Delta {R_m}$ is also representative of the amount of asymmetry in the flow and seems to provide a better collapse of these data.

%Around $\theta \in [-0.13,-0.11]$ in Fig.~\ref{fig4}b, we also notice a region where no dynamo is generated. This regime is characterized by low amplitude magnetic fields, of the order of induced magnetic fields from the ambient field: they are considered non-dynamo but they could also be dynamos with an even lower amplitude that the STAT LOW regime. This type of regimes may also be present at lower and maybe higher forcing. But no experimental point of this type was recorded in the case of Fig.~\ref{fig4}a and c, because of lack of resolution in $\theta$, since this ``no dynamo" region has a very small extension in $\theta$.

As was observed earlier, windows of oscillating dynamo can be present between two stationary modes. Fig.~\ref{fig5}b displays the periods measured in all the time dependent cases of the parameter space, as a function of $\vert\Delta {R_m}\vert$. Using this variable, the evolution of the oscillation periods collapses on a single plot. {Note that the collapse is slightly better when only the points related to a given sign of $\Delta {R_m}$ are plotted.  This traces back to experimental imperfections that slightly break the 
symmetry of the parameter space when $F_1$ and $F_2$ are exchanged.} In the cases of transition from time dependent to STAT HIGH regions, the oscillation period diverges, displaying a $(\vert\Delta {R_m}\vert-\vert\Delta {R_m}^c\vert)^{-1/2}$ behavior, where $\Delta {R_m}^c$ is defined as the value of $\Delta {R_m}$ at the onset of each oscillating regime. On the other hand, when oscillatory regimes approach the STAT LOW region, the transition develops with a finite period.

\section{Low dimensional dynamics}
%%%%%%%%%%%

\subsection{Observations}

In order to understand better the observed transitions, we can display a cut in the phase space for the magnetic field recorded in one location, by representing a component of the field as it evolves in time, versus the same component delayed by a time $\tau$. Fig. \ref{fig6}a shows examples of such trajectories in the phase space for different regimes: two fixed points are observed, corresponding to the STAT HIGH regime (green curve) and to the STAT LOW regime (blue, where a transient regime can also be seen). Then the two limit cycles (magenta and red) correspond respectively to a periodically oscillating regime and a randomly reversing one. Examples of the 4 types of dynamo regimes are shown in Fig.~\ref{fig6}b to e. The transition between the STAT LOW and the STAT HIGH regimes, via the two time dependent regimes, corresponds to the rising branch in Fig. \ref{fig4}c, in the region of $0.2<\theta<0.27$.

\subsection{Comparison with a model}
%%%%%%%%%%%%
Several studies have shown how features of the Earth paleomagnetic records~\cite{Melbourne:Cargese:01,Ryan:EPL:08} or sunspots activity~\cite{Knobloch:MNRAS:96,Mininni:SP:01},  can be described as the dynamics of a low dimensional system. In the VKS experiment, the low dimensional nature of the dynamics of the magnetic field has been emphasized in~\cite{Ravelet:PRL:08}. A model based on the interactions of two magnetic modes having the symmetry of a dipole and of a quadrupole has been introduced in~\cite{Petrelis:JPCM:08}. In this framework two types of bifurcations from stationary to oscillatory dynamo regimes have been introduced. When the stationary state is far enough from the dynamo threshold, the transition to an oscillatory regime occurs via a saddle-node bifurcation~\cite{Petrelis:JPCM:08}. When both instability modes are nearly marginal, {\it i.e.} in the vicinity of a codimension-two bifurcation, bistability of stationary and oscillatory states occurs and the dynamics is more complex~\cite{Berhanu:JFM:09}. We expect that the STAT HIGH and STAT LOW stationary dynamos bifurcate to time dependent regimes in related ways. 

\begin{figure}[h!]
{\begin{minipage}{0.5\textwidth}
\centering
\centerline{\includegraphics[width=8.5cm]{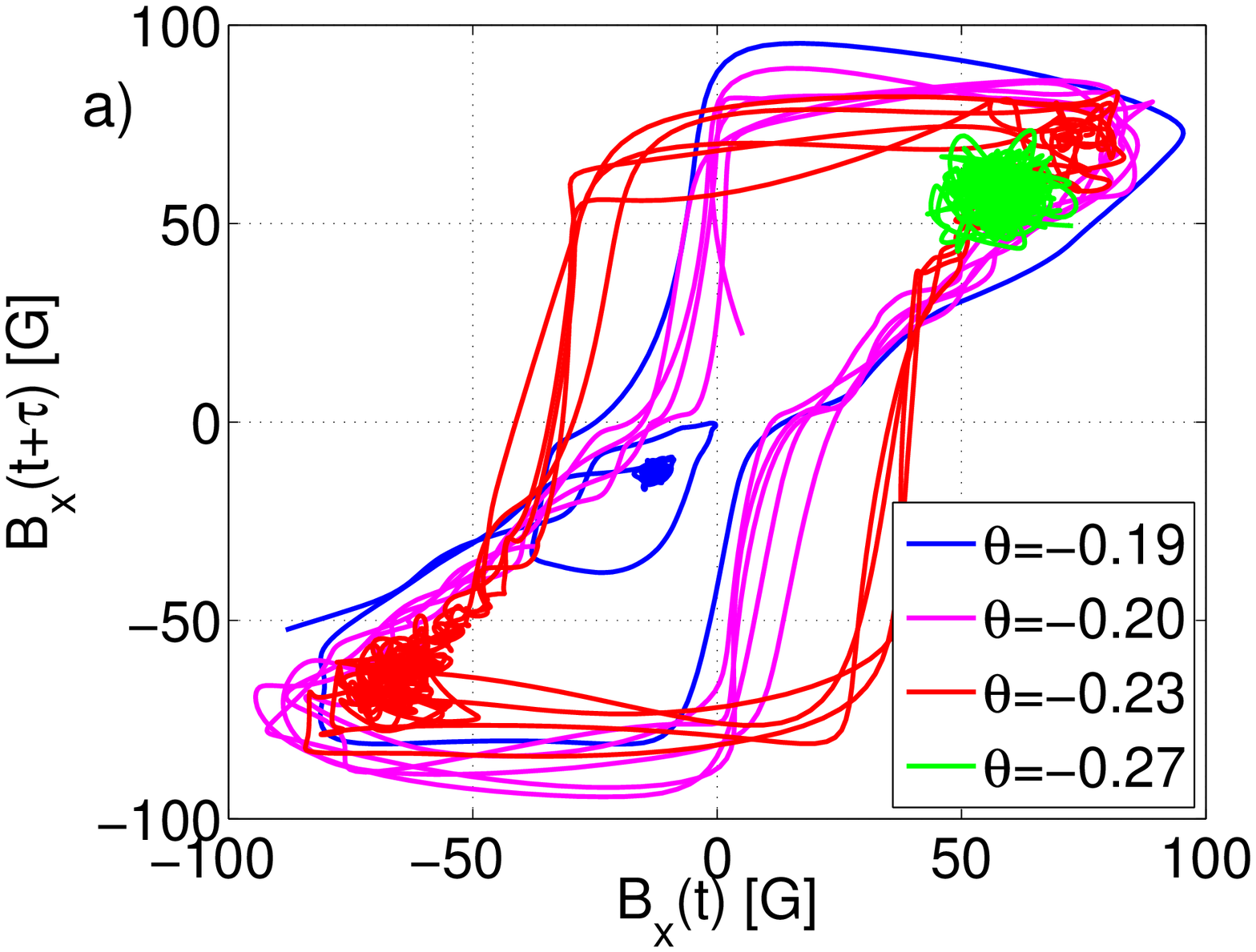}}
%\vspace{2mm}
\includegraphics[width=.46\textwidth]{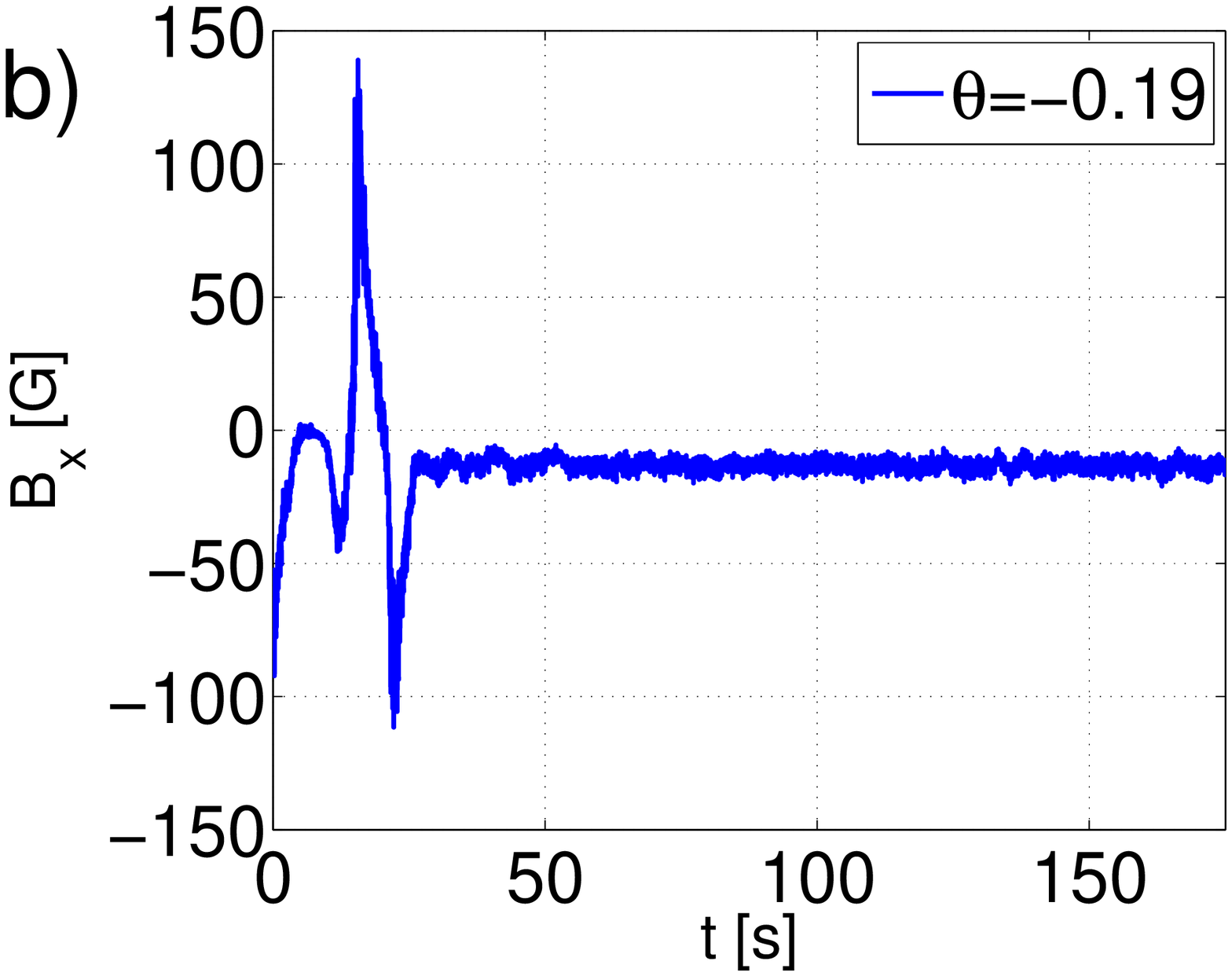}
\includegraphics[width=.46\textwidth]{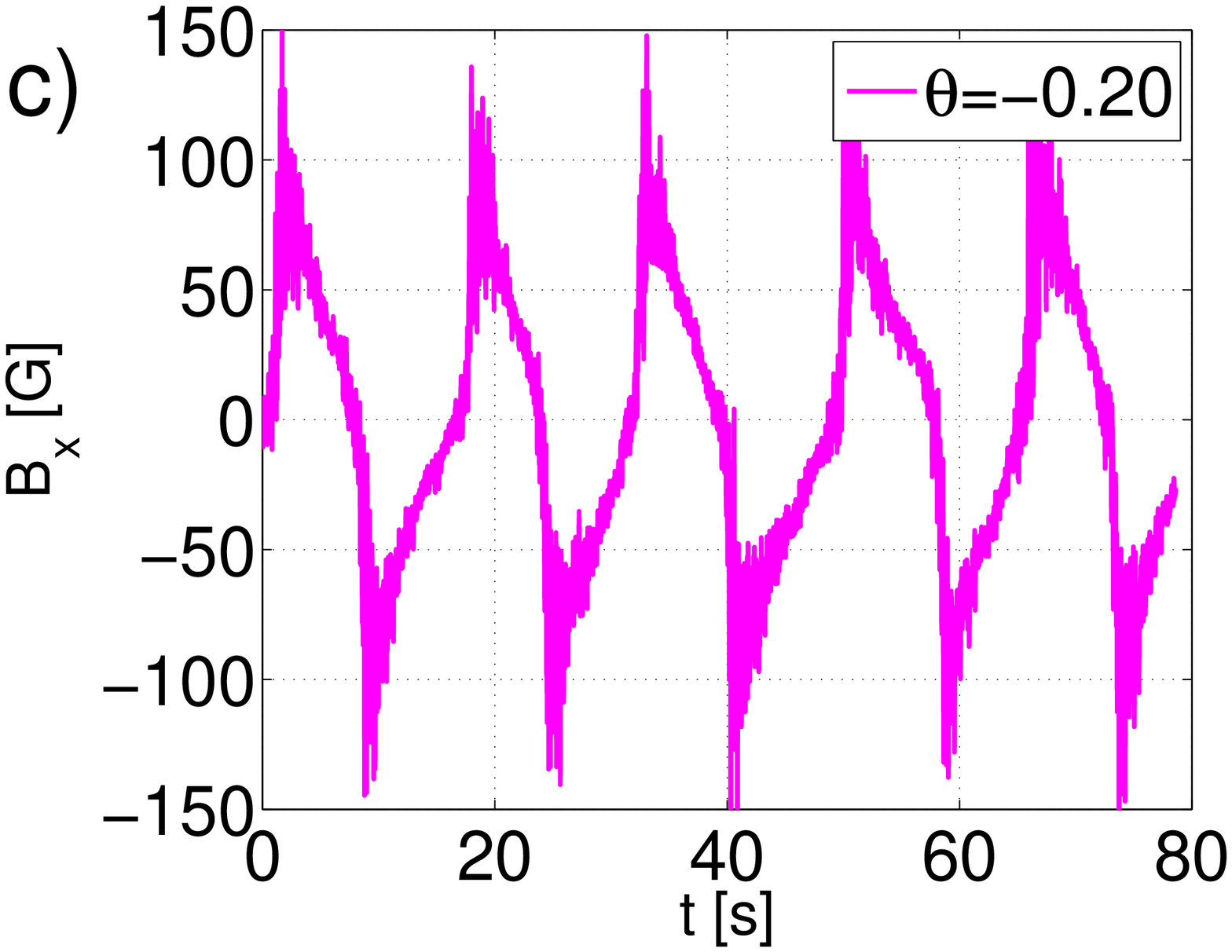}\\
%\vspace{2mm}
\includegraphics[width=.46\textwidth]{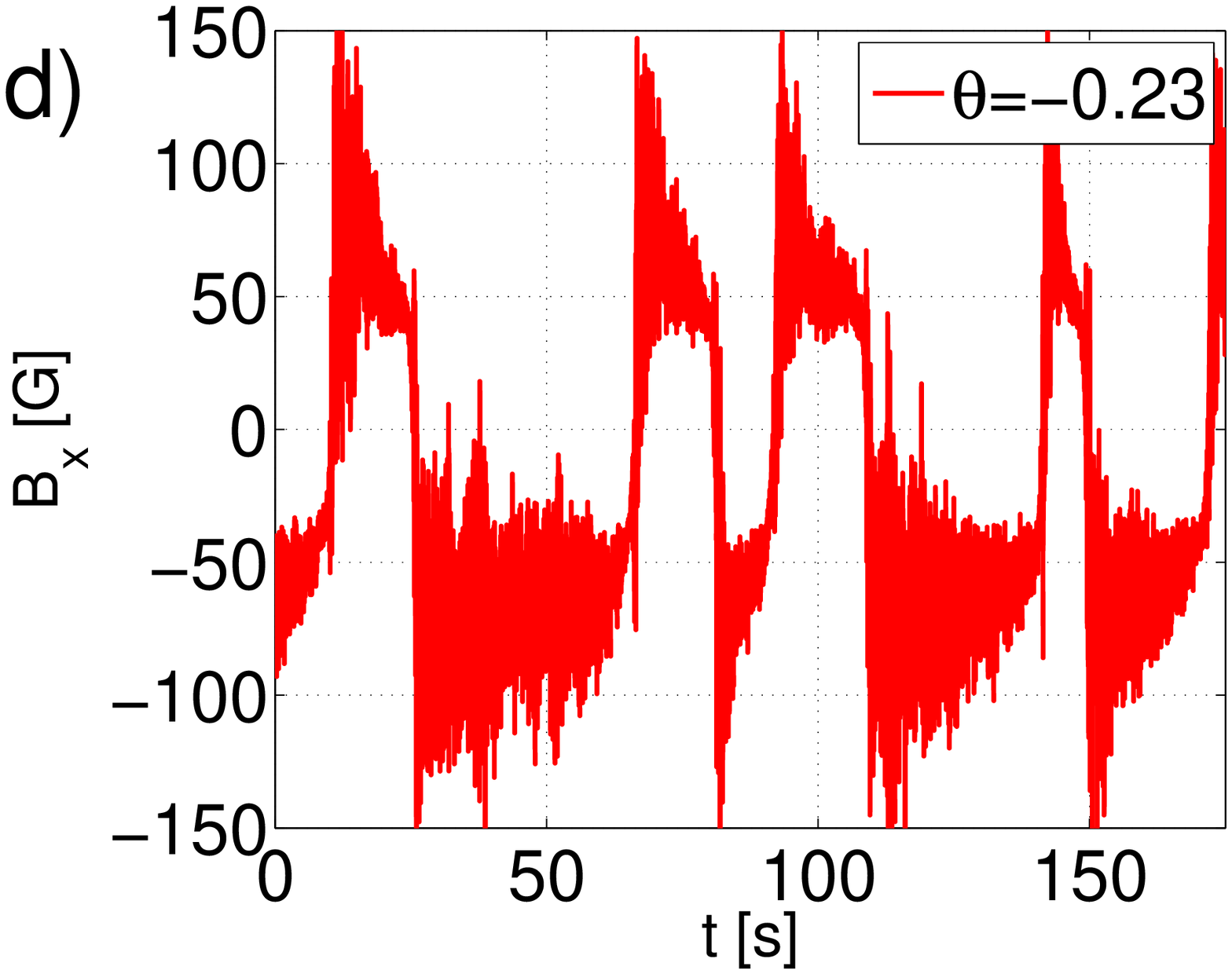}
\includegraphics[width=.46\textwidth]{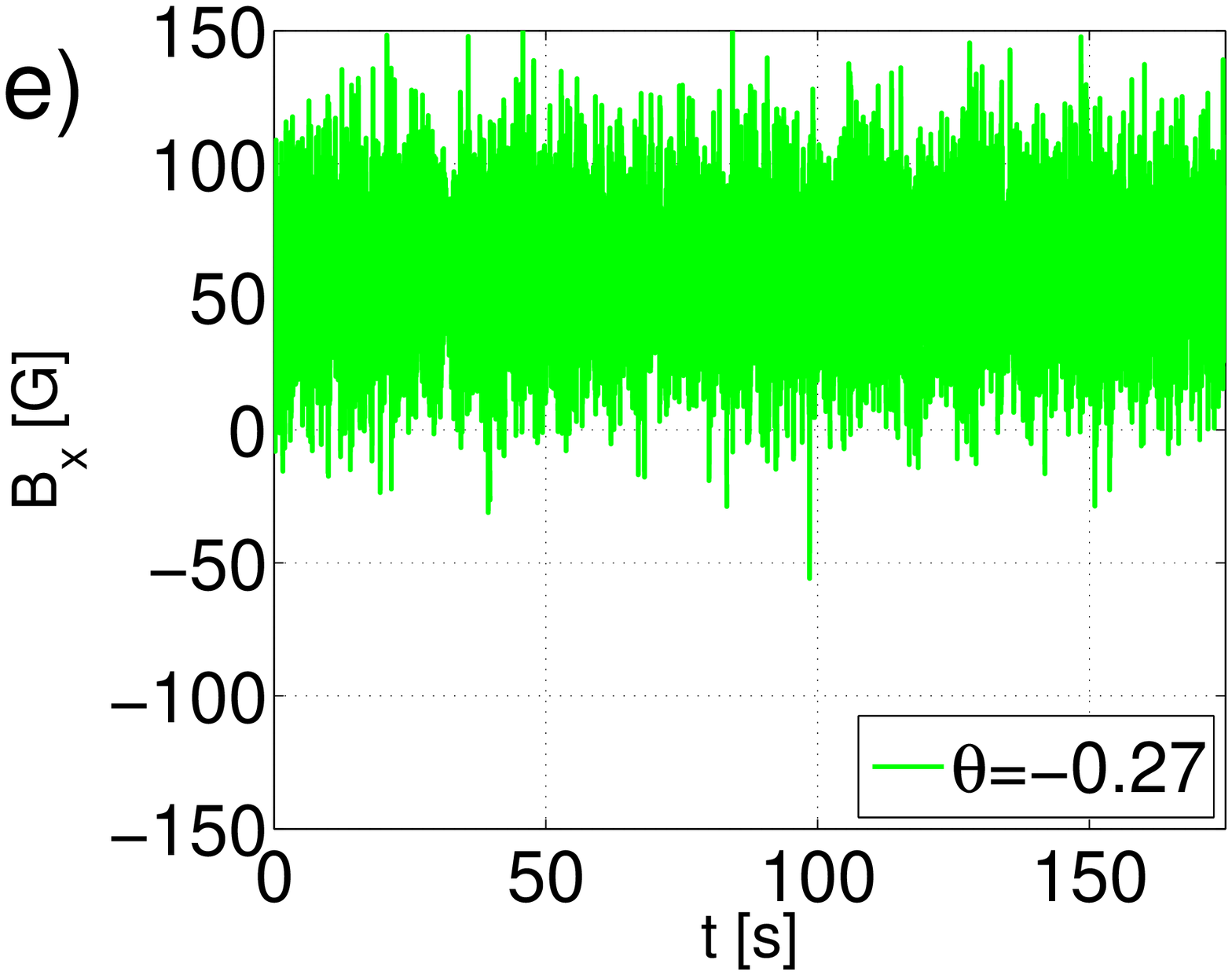}

\caption{a) Phase space for the axial component recorded at the location \#2, for several kinds of dynamos obtained for different values of $\theta$. The time signal were low-pass filtered at 2 Hz and $\tau=1$ s. Examples of the corresponding time signals for: b) the low stationary dynamo with a part of transient regime (blue in the phase space plot) c) a periodic reversing dynamo (magenta), d) a random reversing dynamo (red) and e) a high stationary dynamo (green). For these cases ${R_m}$ is roughly constant around 41.2 ($F_1+F_2=44$ Hz).}
\label{fig6}\end{minipage}}
\end{figure}

The STAT HIGH regime is at finite distance from the dynamo threshold when it undergoes a time dependent instability, and we ascribe this transition to a saddle-node bifurcation. The fixed point related to STAT HIGH collides with an unstable fixed point and a limit cycle is generated (see figure \ref{fig6}a), which connects the system to the opposite polarity. As expected in the vicinity of a saddle-node bifurcation, a divergence of the period of oscillation is observed with a $(\Delta {R_m}-\Delta {R_m}^c)^{-1/2}$ behavior (see figure \ref{fig5}b). Random reversals (figure \ref{fig6}d) appear at the border between the STAT HIGH stationary state (figure \ref{fig6}e) and the periodic regime (figure \ref{fig6}c). Indeed, slightly before the saddle-node bifurcation, small fluctuations are enough to push the system beyond the unstable fixed point and thus generate a field reversal~\cite{Petrelis:PRL:09}. Although fluctuations are necessary to escape from the metastable fixed point STAT HIGH, most of the trajectory that connects this point to its opposite in phase space is driven by the deterministic low dimensional dynamics. This explains why trajectories related to different reversals are robust and can be superimposed~\cite{Monchaux:PoF:09}. But the time between two reversals is random because turbulent fluctuations acting as noise trigger the escape from STAT HIGH. This waiting time is governed by the distance to the saddle-node bifurcation and by the intensity of the fluctuations. It can be very long compared to the duration of a reversal when the distance to the saddle node bifurcation increases because the mean exit time depends exponentially on the system parameters (see equation (6) in~\cite{Petrelis:PRL:09}).

\begin{figure}
\centerline{\includegraphics[width=8.5cm]{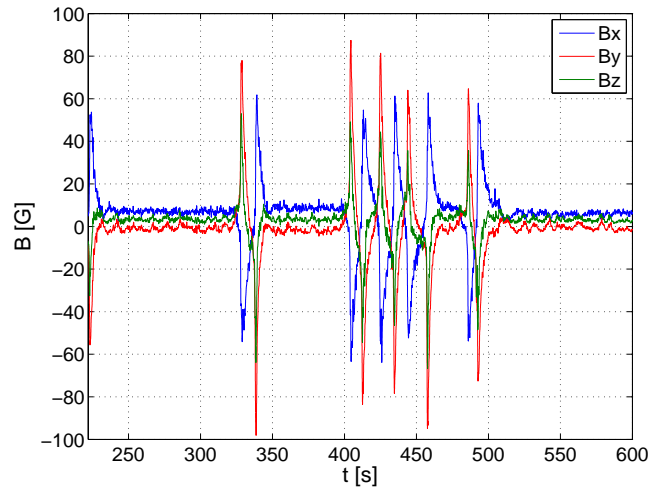}}
\caption{{Example of time signal at $F_1$ = 16.75 Hz and $F_2$ = 25.25 Hz ($\theta=0.2$).}}
\label{fig8}
\end{figure}

We propose that the STAT LOW regime bifurcates to a time dependent regime in a different way, similar to the transitions observed near the regime where only one impeller is rotating, a situation described in details in~\cite{Berhanu:JFM:09}. The transition from STAT LOW occurs in the vicinity of non-dynamo regimes, as shown in figures 1 and 4. Ascribing the bifurcation to the features of a codimension 2 point, one has the existence of two stationary states with opposite polarities which are encircled in phase space by two limit cycles, an unstable one and a stable one. The unstable orbit separates the stationary states and the stable oscillatory one. When $R_m$ is increased from the stationary state in the bistable regime, the unstable limit cycle bifurcates via a double saddle homoclinic connection with zero, generating  two unstable limit cycles located around each stable fixed point. Then, these limit cycles shrink on the fixed points which become unstable. The system jumps to the stable limit cycle with finite oscillation period. When the forcing is decreased from the oscillatory state in the bistable regime, the system stays on the stable limit cycle until it collides with the unstable one and disappears. The magnetic field then jumps to one of the stationary states. This scenario is in agreement with our observations. It also predicts a region of bistability, clearly evidenced for the one-impeller flow (at $\theta$=1)~\cite{Berhanu:JFM:09} and also observed in a narrow range of parameters around $\theta=0.2$, {as can be seen in Fig.~\ref{fig8}, showing a time signal where the dynamo spontaneously jumps from a STAT LOW regime to an oscillatory regime}~\cite{Berhanu:these:08}. 

\section{Discussion}
%%%%%%%%%%%

We first consider several aspects related to dynamo generation and then discuss the modeling of the dynamical regimes. Our results indicate that the iron impellers play a crucial role. As shown in Table 1, no dynamo is generated when driving the flow by impellers made of other materials. In addition, when the flow is driven by an iron impeller and a stainless steel one, only the rotation of the iron impeller gives rise to a dynamo and no further time dependent regime is observed. The exact influence of the iron impellers is still an open question, although it has been emphasized in several studies: modeling~\cite{Petrelis:GAFD:07}, experimental~\cite{Verhille:NJP:10} and numerical~\cite{Giesecke:PRL:10,Gissinger:EPL:09}. New experimental runs, using impellers with disks and blades made of different materials are underway in the VKS set-up, and will hopefully contribute to the understanding of this issue. 

We have also observed that the dynamo threshold can depend on the flow characteristics. For instance, without the inner ring,  the threshold has not been reached when the impellers are rotated in the scooping direction. This observation, together with others mentioned in section~\ref{config}, implies that certain aspects of the flow characteristics play a role in the onset of the instability.

Once the threshold is reached for which a dynamo is generated, we have shown that owing to the asymmetry that can be introduced by rotating both impellers at different frequencies, a rich variety of dynamical regimes arises. 
Experiments in an equivalent water device have been performed in order to check whether the dynamics of the magnetic field is related to instabilities of the flow in the non dynamo regime. They showed that a hysteretic bifurcation is observed at $\vert\theta\vert=0.09$, between a flow with two recirculation cells and a flow with one recirculation cell~\cite{Cortet:PoF:09}. In the same set-up, adding the inner ring in the median plane, this hydrodynamic bifurcation is moved to a larger asymmetry ($\vert\theta\vert=0.16$), showing that the presence or absence of the inner ring can strongly change the flow characteristics.
As to the magnetic bifurcations observed in the VKS experiment, we have seen that at $\vert\theta\vert=0.09$, a transition between high amplitude and low amplitude steady dynamo occurs, sometimes via a time-dependent one. But this feature is robust whether the inner ring is present or not (see figure 18 in~\cite{Monchaux:PoF:09}). Other magnetic bifurcations are present around $\vert\theta\vert=0.16$ with the inner ring and, without inner ring, at higher $\vert\theta\vert$ values (0.2$<\vert\theta\vert<$0.3, dependent on $R_m$) (see Fig.~\ref{fig4}).

We also evidenced other situations where some transitions take place that are not driven by hydrodynamics instabilities: for example, we have observed that a periodically oscillating regime observed at $T=145^\circ$C bifurcates to a chaotically reversing one when the temperature drops to $T=120^\circ$C (see figure 22 in~\cite{Monchaux:PoF:09}). In this case the flow presumably remains unchanged ($\vert\theta\vert=0.16$), while $R_m$ is changed from 42.3 to 44.4 via variations of the sodium electrical conductivity. More generally, one may argue that flows driven at a constant $\vert\theta\vert$, for which bifurcations of magnetic behaviors are observed, as discussed in section 3,  evolve with the same geometrical characteristics. In VK flows, erratic changes in the structure of the flow have however been reported even when the driving (and everything else such as temperature) had been kept to steady values~\cite{Cortet:PoF:09,Torre:PRL:07}. Thus, one can infer that hydrodynamic bifurcations are not necessary for a magnetic bifurcation to occur, but in some cases, a correlation can exist. 

As explained above, the dynamics of the magnetic field reported in this study can be captured by a minimal model which involves the interactions of two magnetic modes. These two modes have the symmetries of an axial dipole and of a quadrupole. The dynamics resulting from the non-linear interactions of two competing modes~\cite{Petrelis:JPCM:08} describes the transitions between neighboring regimes, if one assumes that the two modes are simultaneously marginally stable. Dipole and quadrupole modes have been observed to have nearly the same threshold, in analytical studies of earthlike systems~\cite{Proctor:GAFD:77} as well as in numerical simulations of the Earth dynamo \cite{Christensen:PEPI:01}. In addition, numerical studies aimed at modelling the VKS experiment show that axial dipolar and quadrupolar modes have nearly the same threshold and that their interaction leads to a time dependent regime when the impellers rotate at different rates~\cite{Gissinger:EPL:09}.

In this description using amplitude equations for the dipolar and quadrupolar components of the magnetic field, the fluid parameters, the flow characteristics and the boun\-dary conditions determine the values of the coefficients of the equations. The level of turbulent fluctuations, which changes with $\theta$~\cite{Cortet:PoF:09}, is taken into account through multiplicative noise in the model. However, the deterministic part of the dynamics does not explicitly involve velocity modes.  Besides describing the bifurcations reported in this paper, this two-dimensional phase space of the deterministic dynamics is crucial to explain why the mean value of the magnetic field should vanish in the time periodic regime, and to correctly predict the shape of the reversals~\cite{Petrelis:JPCM:08}.  Including velocity modes will modify the geometry of the phase space and this will be likely to generate behaviors in disagreement with the experimental observations. More generally,  no experimental evidence about the dynamics of the magnetic field requires the inclusion of any additional velocity mode in the framework of this model.

There are of course other issues in the VKS experiment besides the dynamical regimes and transitions described here. For instance, the detailed mechanisms of magnetic field generation and saturation still need to be clarified.  A future study combining informations from torque measurements and local velocity measurements (using either potential measurements~\cite{Ricou:IJHMT:1982} or Doppler velocimetry~\cite{Takeda:NT:87,Nataf:PEPI:08}) could contribute to a better understanding of magnetohydrodynamics features in the VKS dynamo. \\

\noindent {\bf Acknowledgements:} We thank M. Moulin, C. Gasquet, J.-B Luciani, A. Skiara, N. Bonnefoy, D. Courtiade, J.-F. Point, P. Metz and V.  Padilla for their technical assistance. This work is supported by ANR-08-BLAN-0039-02,  Direction des Sciences de la Mati\`ere et Direction de l'\'Energie  Nucl\'eaire of CEA, Minist\`ere de la Recherche and CNRS. The experiment is operated at CEA/Cadarache DEN/DTN.

\bibliographystyle{epj}
\bibliography{Berhanu}% Produces the bibliography viaBibTeX.

\end{document}